%% file: asplos24-paper-template.tex
\documentclass[conference]{IEEEtran}
\ifCLASSINFOpdf
\else
\fi
\usepackage{tikz}

\usepackage[normalem]{ulem}
\usepackage{multirow}
\usepackage{booktabs}
\usepackage{tabularx}
\usepackage{subfigure}
\usepackage{algorithm}
\usepackage{algpseudocode}
\usepackage{amsmath}
\usepackage{amssymb}
\usepackage{threeparttable}
\usepackage{rotating}
\usepackage[utf8]{inputenc}

\begin{document}

\date{}

\title{
\emph{CAT} : A GPU-Accelerated FHE Framework\\ with Its Application to High-Precision Private Dataset Query
}

\author{
{\rm Qirui Li}\\
IDEA Institute
\and
{\rm Rui Zong}\\
IDEA Institute
}
\maketitle

\thispagestyle{empty}

\begin{abstract}

We introduce an open-source\footnote{https://github.com/Rayman96/CAT} GPU-accelerated fully homomorphic encryption (FHE) framework \emph{CAT}, which surpasses existing solutions in functionality and efficiency. \emph{CAT} features a three-layer architecture: a foundation of core math, a bridge of pre-computed elements and combined operations, and an API-accessible layer of FHE operators. It utilizes techniques such as parallel executed operations, well-defined layout patterns of cipher data, kernel fusion/segmentation, and dual GPU pools to enhance the overall execution efficiency. In addition, a memory management mechanism ensures server-side suitability and prevents data leakage.

Based on our framework, we implement three widely used FHE schemes: CKKS, BFV, and BGV. The results show that our implementation on Nvidia 4090 can achieve up to 2173$\times$ speedup over CPU implementation and 1.25$\times$ over state-of-the-art GPU acceleration work for specific operations. What's more, we offer a scenario validation with CKKS-based Privacy Database Queries, achieving a 33$\times$ speedup over its CPU counterpart. All query tasks can handle datasets up to $10^3$ rows on a single GPU within 1 second, using 2-5 GB storage.

Our implementation has undergone extensive stability testing and can be easily deployed on commercial GPUs. We hope that our work will significantly advance the integration of state-of-the-art FHE algorithms into diverse real-world systems by providing a robust, industry-ready, and open-source tool.

\end{abstract}

\section{Introduction}\label{sec::intro}\input{sec1-introduction}
\section{Background and Motivation}\label{sec::backg}\input{sec3-background}

\section{Architecture Design Overview}\label{sec::overview}\input{sec4-architectureoverview}
\section{Application to PDQ}\label{sec::app-PDQ}\input{sec5-application}

\section{Evaluation and Results}\label{sec::eval}\input{sec6-evaluation}

\section{Summary and Future Plans}\label{sec::sum-plan}\input{sec7-summary}

\section*{Acknowledgement}
Projects and people supporting this work will be added later in compliance with the double blind policy.

% This document is revised from submissions guides of ASPLOS'21 to ASPLOS'23. Thank
% you Emery Berger, Christos Kozyrakis, Shan Lu, Thomas Wenisch, Michael Swift, and Natalie Enright Jerger!

\clearpage
\bibliographystyle{plain}
\bibliography{references}

% \clearpage
% \section*{Appendix}\label{sec::appendix}\input{sec8-appendix}

\end{document}

%% file: sec1-introduction.tex
%FHE隐私计算地位和主要场景，以及FHE最主要的功能特点

Fully homomorphic encryption (FHE), first proposed in 1978~\cite{rivest1978data}, enables directly operating on encrypted data while yielding the same encrypted results as if the operations were run on plaintexts.
FHE stands as a powerful tool in privacy computing, offering secure outsourced computation without revealing any data.
In 2009, Gentry~\cite{gentry2009fully} gave the first FHE construction based on the concept of bootstrapping, which homomorphically evaluates the decryption process.
Since then, many innovative works have been proposed, and the most distinguished FHE schemes can be classified into three categories: 1) CKKS\cite{DBLP:conf/asiacrypt/CheonKKS17} for float number arithmetic, 2) BFV~\cite{DBLP:journals/iacr/FanV12} and BGV~\cite{DBLP:journals/toct/BrakerskiGV14} for integer arithmetic, 3) FHEW~\cite{DBLP:conf/eurocrypt/DucasM15} and TFHE~\cite{DBLP:journals/joc/ChillottiGGI20} for Boolean operations. 

%性能问题
While conceptually ingenious, FHE is still far from widespread adoption due to challenges in computational efficiency, data expansion, and functional limitations. 
Even with the most highly optimized CPU-based FHE libraries, computations on encrypted data still cost $10^3\times$ to $10^6\times$ more time and $10^2\times$ to $10^5\times$ more memory than equivalent plaintext computations~\cite{DBLP:conf/hpca/FanWXHMZ23}.
To address these issues, FHE SIMD (Single Instruction Multiple Data) operation~\cite{DBLP:journals/dcc/SmartV14,DBLP:conf/asiacrypt/CheonKKS17} has been proposed, which encrypts multiple plaintexts into a single ciphertext and enables simultaneous evaluation of subsequent operations.
Another way to enhance FHE computational efficiency is through highly parallel hardware-based implementations.
There have been numerous attempts and research efforts based on different types of hardware, \emph{e.g.}, FPGA~\cite{DBLP:conf/asplos/RiaziLPD20,DBLP:conf/hpca/RoyT0VV19,DBLP:conf/reconfig/KimLCCR19,DBLP:conf/dac/RenCGLZLZZWZLCHWNX23,kim2020hardware,DBLP:journals/tc/MiglioreRLTFG18}, GPU~\cite{DBLP:conf/hpca/FanWXHMZ23,alves2021faster,DBLP:journals/tetc/BadawiPAVR21,9481143,al2020privft,al2018high,kim2020accelerating,jung2021over,DBLP:conf/asap/LeeLCP15,morshed2020cpu}, and ASIC~\cite{SHARP,ARK,BTS,BASALISC,F1,Craterlake,REED}. 
Most FPGA-based works focus on accelerating specific underlying mathematical operations or HE operations within the algorithm. 
ASIC-based algorithm implementations currently offer the highest performance and support bootstrapping operations, but their high cost and limited flexibility hinder real-world adoption.
GPU-based implementations face on-chip memory limitations that restrict kernel stacking for certain functions.

In this paper, we present an all-in-one GPU-accelerated FHE implementation spanning from fundamental hardware realization to concrete application scenarios. We refer to our framework as \emph{CAT}, an acronym for Cipher-Acceleration-Textile. This nomenclature reflects its exquisite proficiency in overseeing the entire lifespan of accelerated FHE workloads. Within our framework, cipher data seamlessly flows akin to shuttles and threads in a textile machine. Our key contributions are threefold:

\begin{enumerate}
    \item[1)] \textbf{Three-Layer Framework}

We propose a three-layer framework for GPU-based FHE implementation, where the foundational layer provides core mathematical operations; the intermediate layer bridges with pre-computed elements and combined operations; and the topmost layer exposes FHE operations, such as encryption and multiplication, as API interfaces to users. 
To enhance efficiency, scalability, and security across all layers, we equip them with innovative techniques: parallel executions, well-defined data layouts, kernel fusion/segmentation. And we introduce two GPU pools, the Stream Pool leverages these techniques to maximize performance, while the Memory Pool ensures memory usage efficiency for server-side suitability.

    \item[2)] \textbf{High Performance Implementation} 
    
    Unveiling the power of our three-layer framework, we successfully implement three widely-used FHE schemes (CKKS, BFV, BGV) with remarkable performance improvements.
    To validate their practicality, we further deploy these implementations on several Nvidia commercial GPUs, demonstrating their potential across different GPU architectures.
    On Nvidia 4090, our implementations achieve astounding speedups of 2173$\times$ over the CPU-based SEAL library and 1.25$\times$ over the state-of-the-art public GPU implementations for specific operations;
    % on Nvidia 3090, our implementations achieve astounding speedups of 844$\times$ over the CPU-based SEAL library and and \textcolor{red}{???}$\times$ over other public GPU implementations for specific operations;

    \item[3)] \textbf{Scenario Validation} 
    
    To demonstrate its real-world applicability, we implemented a private database query protocol based on our high-performance CKKS implementation. This protocol achieves functional completeness, high precision, and high efficiency:
    
    \textbf{$\cdot$} Supports both search and computation in a single query;
    
    \textbf{$\cdot$} Achieves 32-bit precision for non-linear operations;
    
    \textbf{$\cdot$} Exhibits a 1-second query latency on $10^3$ rows with 2-5GB of storage.

\end{enumerate}

In Section~\ref{sec::backg}, we discuss the background knowledge and our motivations,
In Section~\ref{sec::overview}, we give a detailed introduction of our framework,
In Section~\ref{sec::app-PDQ}, we introduce our PDQ protocol,
In Section~\ref{sec::eval}, we give the implementation results,
In Section~\ref{sec::sum-plan}, we summarize our work and list future plans. 

%% file: sec3-background.tex
\subsection{Notations}
This section introduces basic notations and HE-related terminologies.
Raw data is called \emph{clear text} and it is encoded to \emph{plaintext}. The data is converted into \emph{ciphertext} via encryption.

RLWE is an encryption methodology widely used in HE schemes, which is the ring version of LWE-based cryptosystems. 

The plaintext space is determined by an integer parameter $\emph{t}$, a plaintext is an element of $\mathcal{R}_t=\mathcal{R}/(t\mathcal{R})$, \emph{i.e.} a polynomial of degree at most $n-1$ with coefficients in $\mathbb{Z}_t$. Ciphertexts are elements of $\mathcal{R}_q=\mathcal{R}/(q\mathcal{R})$, \emph{i.e.} polynomials of degree at most $n-1$ with coefficients in $\mathbb{Z}_q$.
The parameters $q$ and $n$ determines the security of the underlying RLWE cryptosystem.  
For more information, refer to~\cite{albrecht2021homomorphic}.

\subsection{RNS Representation}
\label{sub-rns}
Usually, $q>>t$, the length of $q$ can be hundreds of bits. Directly computing on these big numbers is inefficiency.
$q$ can be a product of small pairwise coprime numbers $q = \prod^k_{i=1} q_i$.
Then, the HE operations can be implemented as a Residue Number System (RNS), using
the Chinese Remainder Theorem (CRT) which essentially offers a ring isomorphism $\mathbb{Z}_q \rightarrow\prod^k_{i=1} Z_{q_i}$, to manipulate
the large coefficients of ciphertext polynomials, as the isomophism can be extended to polynomials: 
$\mathcal{R}_q \simeq \mathcal{R}_{q_1}\times\mathcal{R}_{q_2}\times\cdots \times\mathcal{R}_{q_k} $.

The RNS concept is also well-suited for coefficient level manipulation in HE operations. 
The level concept is important for operations such as Rescale in CKKS and ModulusSwitch in BFV/BGV.

\subsection{FHE Hardware Accelration}
Research on hardware acceleration for FHE primarily focuses on three approaches: FPGA, ASIC, and GPU.

Many FPGA-based FHE implementations focus on optimizing some specific operations, such as the underlying mathematical operation NTT~\cite{kim2020hardware,DBLP:journals/tc/MiglioreRLTFG18} or HE operators such as key-switch~\cite{DBLP:conf/asplos/RiaziLPD20} and moduluar multiplication~\cite{DBLP:conf/reconfig/KimLCCR19} . 
In~\cite{DBLP:conf/hpca/RoyT0VV19}, the authors give an Arm-FPGA co-processor framework to achieve $13\times$ speedup of BFV compared with the CPU implementation. 
In~\cite{DBLP:conf/dac/RenCGLZLZZWZLCHWNX23}, an HE accelerator for high-performance matrix-vector product is given and validated in multiplication-centric scenarios.

Compared to software implementations, ASIC-based implementations can achieve up to 10000$\times$ speedup and thus become the most notable acceleration results. However, whether these efforts can truly advance the practical deployment of FHE requires further validation and optimization. 
For example, due to the integrated-into-single-chip design, the chip area of ~\cite{SHARP,ARK,BTS,BASALISC,F1,Craterlake} is large, which can be up to more than 100 mm$^2$.
In~\cite{REED}, the authors propose the first chiplet-based FHE accelerator to solve the scalability problem. 
However the development cycle, market readiness, and costs need further clarification and optimization, especially considering the high development cost and extended time-to-market.

Many GPU-based studies also only focus on accelerating basic operations~\cite{DBLP:conf/hpca/FanWXHMZ23,alves2021faster,DBLP:journals/tetc/BadawiPAVR21,9481143,al2020privft,al2018high,kim2020accelerating}, and excluding bootstrapping acceleration. The work~\cite{jung2021over} firstly accelerates all HE operations, including CKKS bootstrapping, and overcomes the GPU's off-chip memory bottleneck through kernel fusion, which achieve a bootstrapping process that is 242 times faster than on a CPU. However, the GPU's on-chip memory is still the major limitation restricting kernel fusion for certain functions~\cite{kim2020accelerating}. Also, there are efforts~\cite{DBLP:conf/asap/LeeLCP15,morshed2020cpu} targeting boolean HE schemes.

\subsection{FHE-based Private Database Query}\label{subsec-pdq}

\begin{figure}[!h]
    \centering
    \includegraphics[width = 0.4\textwidth]{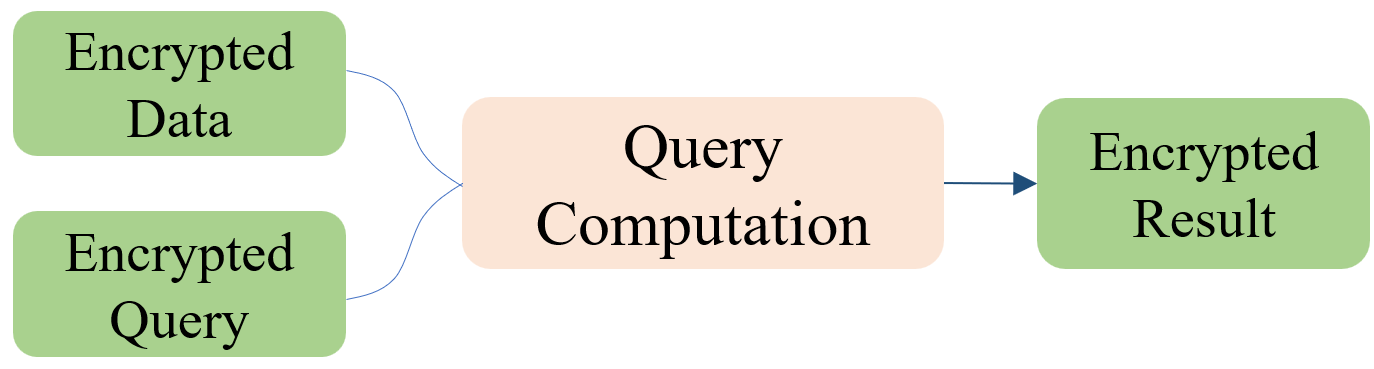}
    \caption{Workflow for a PDQ Query}
    \label{fig:query-overview}
\end{figure}
Private Database Query (PDQ) empowers secure encrypted database queries, safeguarding query confidentiality. It establishes a secure client-server protocol, enabling search and computation processes—including comparison, linear, and nonlinear operations-directly on encrypted data, as shown in Fig~\ref{fig:query-overview}.
FHE offers an elegant approach for PDQ implementation. 
However, its noise budget faces significant depletion due to the inherently complex nature of encrypted comparison operations. This constraint has impeded advanced computations in previous works~\cite{DBLP:journals/tdsc/TanLWRA21/comparison,DBLP:journals/tifs/CheonKK16,DBLP:journals/popets/IliashenkoZ21,DBLP:conf/aina/NarumanchiGEG17/comparison,DBLP:journals/access/KimLLRTW19,DBLP:journals/tdsc/KimLLW18,DBLP:conf/fc/CheonKK15} , as exemplified by Tan et al.~\cite{DBLP:journals/tdsc/TanLWRA21/comparison}, where a single comparison of 64-bit integers necessitates 10 seconds and exhausts the available noise budget, rendering bootstrapping indispensable for further operations.

Lee et al.~\cite{lee2023heaan} prioritized functional completeness and accuracy over efficiency, implementing PDQ using CKKS with bootstrapping. However, unlike BGV/BFV/TFHE, CKKS bootstrapping merely enables further computations without reducing ciphertext error. To compensate, they leveraged the NTL library for enhanced precision, albeit sacrificing performance compared to RNS-based methods~\cite{DBLP:conf/sacrypt/CheonHKKS18}. Notably, a single bootstrapping operation in their implementation takes approximately 25 seconds.

Scheme-switching~\cite{PEGASUS,CHIMERA} offers linear operations or computations up to a certain depth using SIMD-capable algorithms, but it necessitates ciphertext conversion to a different FHE algorithm with efficient bootstrapping when needed. However, these bootstrapping-efficient algorithms typically support narrow data bit-widths. In PDQ operations, ciphertext decomposition during intermediate stages is computationally expensive. Thus, fresh CKKS ciphertexts would also be constrained to limited width data, conflicting with PDQ requirements, rendering scheme-switching unsuitable for PDQ scenarios.

To date, no known work has achieved high efficiency for FHE-based PDQ while also satisfying functional completeness and accuracy requirements, e.g., completing a query under 1 second.

%% file: sec4-architectureoverview.tex
\begin{figure*}[htbp]
\centering
        \includegraphics[scale=0.54]{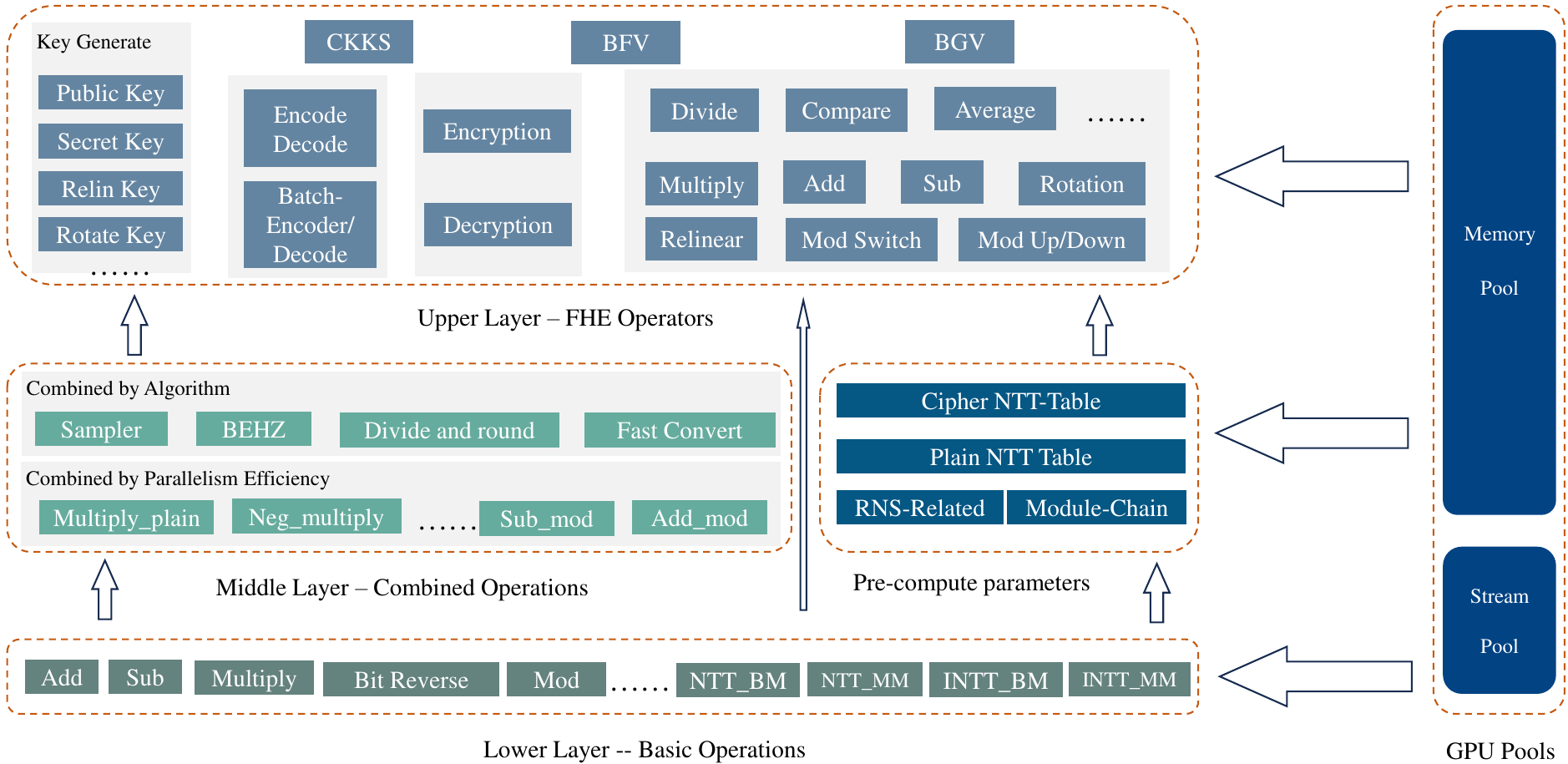}
\caption{Overall architecture of the GPU-accelerated FHE framework}
\label{overall_architecture}
\end{figure*}
% The overall architecture of our framework is shown in Fig \ref{overall_architecture}. The whole framework are designed as leveled structure except for two GPU pools. 

% The lower layer contains all the basic mathematics calculation which are the foundations for the whole framework, like numerical add/sub/multiply and some basic operation in FHE like NTT(Number Theoretic Transform) and bit-reverse. We dedicated to optimize each of the lower operators to ensure the overall performance.

% The middle layer are composed of all the combined operations or the pre-compute parameters, all the operations in this layer will contribute most for the upper layer. There are two types of combined operations, one of which are combined by the theoretical logic, like the BEHZ-algorithm is a method to xxxxxx, making some of lower operators combined together naturally. On the other hand, some of operations in the middle layer are combined just for parallelism efficiency, like 'neg\_multiply' operation logically performs two separate calculation negative and multiply. The two stages can be squeezed into one in order to reduce the latency of starting two GPU kernels to only one time.
% \subsection{Overall architecture design}

Our framework adopts a three-tier architecture, as shown in Figure \ref{overall_architecture}, with two GPU pools providing full-process support.

The bedrock of the framework resides in the lower layer, which houses meticulously optimized fundamental mathematical computations. These include numerical operations such as addition, subtraction, and multiplication, as well as essential FHE operations like Number Theoretic Transform (NTT) and bit-reversal. This optimized foundation ensures robust performance across the entire framework.

% The middle layer consists of combined operations or pre-computed parameters that contribute to the upper layers. There are two types of operations in this layer. The first type results from theoretical logic, such as multiplication under BFV using BEHZ algorithm. These operations naturally combine lower-level operators. The second type combines operations for parallelism efficiency. For example, 'neg\_multiply' performs negative and multiply as a single operation instead of two, reducing latency from launching two GPU kernels to just one.

In the middle layer, combined operations and pre-compute elements, meticulously orchestrate a symphony of efficiency for the upper layers. Two distinct types of combination strategies reside within this layer:

\textbf{Theoretical Concoctions}: Operations such as BFV's multiplication using the BEHZ algorithm\cite{al2019implementation} emerge as natural amalgamations of lower-level operators, directly stemming from theoretical underpinnings.

\textbf{Parallelism Prowess}: Operations like 'neg\_multiply' fuse multiple actions into a single optimized kernel. This ingenious fusion slashes latency by halving the number of GPU kernel launches, unlocking a seamless flow of computations.

The upper layer of the framework is the culmination of the framework's design, providing a comprehensive set of APIs for users. Drawing primarily upon the middle layer's operations and selectively tapping into lower-level functions. Not only focusing on the calculation stage, the accelerated APIs span the entire life cycle of FHE programs: GPU-powered fundamental FHE operations unleash a torrent of computational power; the encode/decode and encrypt/decrypt stages undergo meticulous optimization, ensuring a streamlined data flow; key generation periods of various types benefit from efficiency enhancements, paving the way for swifter cryptographic setups.

% Operators in the upper layer primarily serve as APIs for users, utilizing functions mostly from the middle layer and partly from the lower layer. We have developed APIs for all stages of an FHE program life cycle across the three most widely used FHE schemes. Not only have we accelerated basic calculation APIs using GPUs, but we have also optimized the encode(decode) stage, encrypt(decrypt) stage, and various types of key generation periods.

In addition to the three-tier structure, two crucial GPU pools support all layers: a memory pool that manages the distribution and retrieval of GPU memory (discussed in \ref{subsec-pool}), and a GPU stream pool containing a fixed number of CUDA streams constructed at the beginning of an FHE workload. We design the operators to leverage multi-stream pipelines to reduce running time, detailed in \ref{multi_stream}.

In the following subsections, we will delve into the details of the architecture of the proposed framework. The chapter is organized as follows: Subsection \ref{three_layer} introduces the details of the main body of the three-layer framework, Subsection \ref{subsec-pool} presents the design of the GPU memory pool, and Section \ref{sec::app-PDQ} briefly describes how a high-precision PDQ process runs in our framework.

\subsection{A three leveled acceleration framework}  
\label{three_layer}
We have provided a brief overview of the three-level acceleration framework, describing its theoretical design and composition. In this section, we will delve into key points that contribute to the efficient functionality of this layered structure, particularly in the context of a PDQ process.

\subsubsection{Convert Operator to SIMD Paradigm}
\label{convert_simd}
FHE, grounded in the Residue Number System (RNS) and the Chinese Remainder Theorem (CRT) introduced in \ref{sub-rns}, simplifies the transformation of the calculations into the Single Instruction Multiple Data (SIMD) paradigm. The core concept involves allocating the computation of each integer to an individual GPU thread, thereby converting most operators into element-wise kernels. This approach enhances parallelism and accelerates the processing of cryptographic operations.

The detailed implementation of over 200 kernels, constituting the operators, can be explored in our open-source code. Here, we would like to spotlight several key innovations.

\subsubsection{Multi Level Memory Usage on GPU}

There are three widely-used levels of memory on GPU, namely global memory, shared memory, and register, ordered by their size in descending way and I/O speed in ascending order. Typically, the data we process is stored in the largest global memory most of the time, achieving remarkable acceleration already. To further enhance the performance, we optimize the utilization of shared memory and registers. Leveraging registers is facilitated by declaring const variables in a kernel when they are used multiple times. This approach allows the values to be extracted from global memory to registers, significantly speeding up the retrieval stage. Utilizing shared memory follows a similar principle to prefetching but enables the storage of a larger amount of data at one time. In the following sections, we will present two typical examples illustrating the use of different levels of memory: the calculation of modulo for large integers and the Number Theoretic Transform (NTT).

% As mentioned in Section \ref{sec::backg}, most calculations under FHE end with a mod operation, which can be computationally expensive for large numbers. The \texttt{\%} operator for large numbers in the CUDA toolkit comes with the Barrett algorithm. We have implemented a GPU version of the algorithm demonstrated in Algorithm \ref{module_gpu} \cite{will2014computing}. The algorithm primarily involves dividing the dividend $X$ into a certain number of segments based on the length of the divisor $Y$. Subsequently, the divisor $Y$ is also segmented based on a certain length. Each segment obtained is processed by looking up a precomputed table. The final result is obtained through a sequence of addition, subtraction, and bit-shift operations. The key optimizations for our GPU implementation are as follows: 1. Transfer the contents of the table from global memory to shared memory to accelerate access speed. 2. Optimize for multiple modulus calculations by having each GPU thread compute multiple numbers, reducing the GPU load for table lookup. 3. Optimize for specific arithmetic operations, minimizing arithmetic instruction overhead. 4. Explicitly declare certain cases during table lookup to enhance efficiency.

\textbf{Mod for Big Integers on GPU}

A significant portion of computations in FHE concludes with a mod operation, which can pose computational challenges, especially for large numbers. The \texttt{\%} operator for large numbers in the CUDA toolkit comes with the Barrett algorithm. We have implemented a faster GPU version of the algorithm introduced in \cite{will2014computing}, demonstrated in Algorithm \ref{module_gpu}. The algorithm revolves around segmenting the dividend $X$ based on the length of the divisor $Y$ and, in parallel, segmenting the divisor $Y$ itself. Each segment is then processed by referencing a precomputed table and is prefetched into shared memory, with the final result derived through a series of addition, subtraction, and bit-shift operations. Key optimizations have been integrated into our GPU implementation: 
\begin{enumerate}
    \item[1)]Memory Transfer Optimization: The precomputed table contents are prefetched from global memory to shared memory, boosting access speed, with fully used registers. 

\item[2)]Batch Modulus Calculations: Multiple modulus calculations are optimized by having each GPU thread compute several numbers, effectively reducing the GPU load for table lookup.

\item[3)]Arithmetic Operation Optimization: Specific arithmetic operations are optimized to minimize arithmetic instruction overhead. 

\item[4)]Conditional Declaration: Certain cases are explicitly declared during table lookup, enhancing overall computational efficiency.
\end{enumerate}

% \begin{algorithm}[h]
% \caption{Calculation of modulus for two big integers on GPU using method in \cite{will2014computing}}
% \label{mod_short}
% \begin{algorithmic}[1]

% \Procedure{Modulus}{}
%     \State \textbf{Input:} X, Y
%     \State \quad \textbf{Calculate} $shift_{1st}, shift_{2ed}, M\_2\_S$
%     \State \quad \textbf{Build} Shared Table T\_Shared
%     \State \quad \textbf{Load} pre-compute elements to T\_Shared
%     \State \textbf{Sync threads}

%     \hspace*{\fill}

%     \State $i_{init} \gets \text{GPU thread index} * P$
    
%     \For{$i \gets i_{\text{init}}$ \textbf{to} $i_{\text{init}} + P$}
%         \State $\text{numX} \gets X[i]$

%         \State $\text{T} \gets \text{numX} \gg \text{shift}_{1st} $
%         \State \textbf{Calculating} T with numX, Y, $\text{T\_Shared}$        
%         \State $\text{result[}i\text{]} \gets (\text{T} \ge Y) \, ? \, (\text{T} - Y) \, : \, \text{T}$
%     \EndFor

%     \State \textbf{Output:} result
% \EndProcedure
% \end{algorithmic}
% \end{algorithm}

\begin{algorithm}[!ht]
\caption{Calculation of modulus for two big integers on GPU using method in \cite{will2014computing}}
\label{module_gpu}
\begin{algorithmic}[1]
\Procedure{Set\_higher\_bits\_0(T)}{}
    \State $\text{ploy\_mod\_Y} \gets (2^{len_Y} \mod \text{Y})$
    \While{$(\text{T} \ge (\text{1} \ll len_Y)) $}
        \State $\text{T} \gets \text{T} - (\text{1} \ll len_Y)  + \text{ploy\_mod\_Y} $
    \EndWhile
\EndProcedure

\hspace*{\fill}

\Procedure{Modulus}{}
    \State $\text{shift}_{1st} \gets 64 \div len_Y \ast len_Y$
    \State $\text{shift}_{2ed} \gets ((len_Y - 1) \mod len_{look\_up}) + 1$
    
    \State $\text{idx\_shift}_{1st} \gets len_Y - \text{shift}_{2st}$
    \State $\text{idx\_shift}_{2ed} \gets len_Y - len_{look\_up}$
    
    \State $\text{MAX\_2ED\_SHIFT} \gets \lceil len_Y \mod  len_{look\_up} \rceil$
    \State $\text{SHARED\_SIZE} == 2^{len_{look\_up}}$

    \hspace*{\fill}

    \State $i_{\text{init}} \gets (\text{blockIdx.x} \times \text{blockDim.x} + \text{threadIdx.x}) \times P$

    \State \textbf{shared} $\text{uint64\_t}$ \text{shared\_table[SHARED\_SIZE]}
    
    \If{$\text{threadIdx.x} < \text{table\_size}$}
        \State \text{shared\_table[threadIdx.x]} $\gets$ \text{table[threadIdx.x]}
    \EndIf
    \State \textbf{sync threads}

    \hspace*{\fill}
    
    \For{$i \gets i_{\text{init}}$ \textbf{to} $i_{\text{init}} + P$ \textbf{and} $i < \text{size}$}
        \State $\text{numX} \gets X[i]$
        \State $\text{T} \gets \text{numX} \gg \text{shift}_{1st} $

        \For{$\text{shift} \gets \text{shift}_{1st} - len_Y$ \textbf{down to} $0$ \textbf{by} $len_Y$}
            \State $\text{next} \gets (\text{numX} \gg \text{shift}) \& \text{digits}$

            \State $\text{idx} \gets \text{T} \gg \text{idx\_shift}_{1st} $
            \State $\text{T} \gets \text{T} \ll \text{shift}_{2ed}$
            
            \If{$\text{idx} \neq 0$}
                \State $\text{T} \gets \text{T} + \text{shared\_table[T]}$
                \State $\text{Set\_higher\_bits\_0(T)}$
            \EndIf

            \For{$\text{j} \gets 1$ \textbf{up to} $\text{MAX\_2ED\_SHIFT}$ \textbf{by} $1$}        
                \State $\text{idx} \gets \text{T} \gg \text{idx\_shift}_{2ed}$

                \State $\text{T} \ll len_{look\_up}$
                \State $\text{T} \gets T \& \text{digits} + \text{shared\_table[idx]}$

                \State $\text{Set\_higher\_bits\_0(T)}$

            \EndFor

            \State $\text{T} \gets \text{T} + \text{next}$

            \State $\text{Set\_higher\_bits\_0(T)}$
        \EndFor
        
        \State $\text{result[i]} \gets (\text{T} \ge Y) \, ? \, (\text{T} - Y) \, : \, \text{T}$
    \EndFor
\EndProcedure
\end{algorithmic}
\end{algorithm}

\textbf{NTT\&INTT}

The NTT operation stands out as a pivotal component in FHE, and various endeavors aim to enhance its computational efficiency. A prominent approach involve the utilization of the butterfly method (BM) and the exploration of strategies to eliminate certain intermediate results or reduce the number of calculation steps. In contrast, approaches such as TensorFHE adhere to the original definition of NTT, emphasizing the maximization of matrix multiplication (MM) capabilities on GPUs.  The BM version offers a more generalized solution, requiring computation in $O(\log n)$ steps, yet it might not fully exploit the GPU's inherent computational power. Conversely, the MM version has the advantage of leveraging the giant amount of computation cores of GPU's architecture, achieving considerable efficiency even with computational complexity of $O(n^2)$.

To create a framework that exhibits greater versatility across various GPU architectures, we have implemented both types of NTT methods, denoted as NTT\_BM (butterfly method version) and NTT\_MM (matrix multiplication version), along with their respective inverse counterparts. The implementation of NTT\_MM on standard CUDA cores involves efficient matrix multiplication utilizing shared memory and registers. We employ a trick that when the degree of the polynomial is less than 1024, NTT\_MM is used, and NTT\_BM is chosen otherwise. This adaptive strategy ensures optimal performance based on the specific characteristics of the polynomial being processed.

\subsubsection{Kernel Fusion and Segmentation}
\label{kernel_fusion}
In the middle and upper levels, kernels or operators from lower levels are amalgamated to complete the desired function. Although a straightforward approach involves initiating kernels sequentially, there exist opportunities for enhancements to further elevate overall performance by strategically fusing or segmenting kernels in specific situations. Such optimizations contribute to a more streamlined and efficient execution of the computational tasks at higher levels of the framework.

\textbf{Kernel Fusion}

Kernel fusion merges multiple kernels into one, thus reducing intermediate results, kernel launches, and re-computation.

Take multiplication in CKKS as an example, the multiply of data $X$ and $Y$ involves five stages:
four calculating combinations of $(X_0, X_1)$ and $(Y_0, Y_1)$, and one adding $X_0Y_1$ and $X_1Y_0$ together. While executing these stages separately requires storage for intermediate results, kernel fusion can optimize this process. We can fuse the stages of generating $X_0Y_1$ , $X_1Y_0$ and their addition into one,  directly storing $X_0Y_1+X_1Y_0$ in its destination. Even more, we can fuse all five stages, reducing kernel launches to $1/5$ and eliminating extra memory needs.

\textbf{Kernel Segmentation}

While kernel fusion excels at streamlining short, intensive computations, kernel segmentation shines in scenarios with extended multi-stream workflows. In such cases, synchronous calculation across multiple streams using smaller kernels can outperform a single, massive kernel launch.

In operations like NTT and INTT, which require processing data on each modulus, segmentation often outperforms a single large kernel. While a monolithic kernel might reduce launch overhead, internal linear execution can hinder performance. Instead, breaking the kernel into modulus-specific cases, assigning them to separate streams, and launching them concurrently often leads to remarkable efficiency gains. Section \ref{multi_stream} delves further into multi-stream usage and stream pools.

\textbf{Trade off Between Fusion and Segmentation}

The optimal choice between kernel fusion and segmentation hinges on a delicate trade-off. Kernel fusion outperforms when multiple kernels share redundant calculations and intermediate results, effectively streamlining the process. Conversely, kernel segmentation excels when individual kernels exhibit low register requirements, lengthy execution times, and parallelizable tasks across distinct data segments, enabling concurrent execution for enhanced performance.

\subsubsection{Date layout and Indexing}

\begin{figure}[h]
{
        \centering
        \includegraphics[scale=0.15]{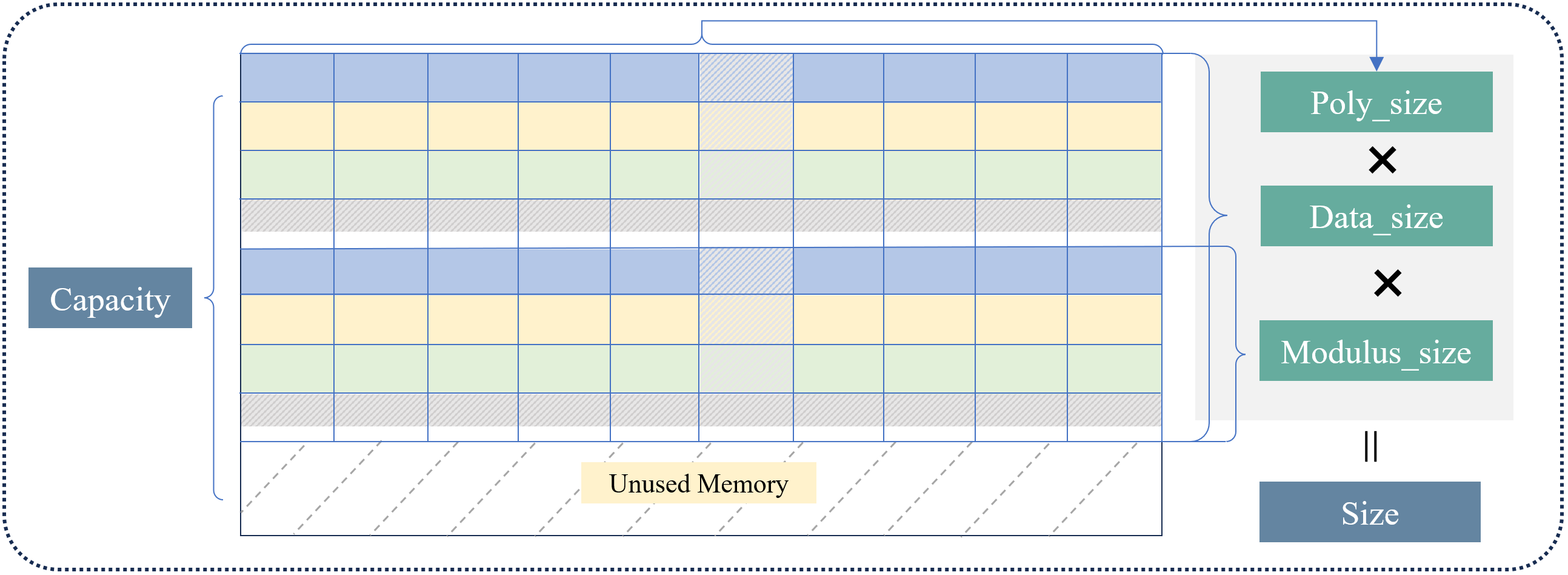}
}
\caption{Structure for storing underlying data}
\label{data_structure}
\end{figure}

To address the performance costs associated with data transfers and non-continuous indexing, we've designed the $C\_Data$ structure (Fig.\ref{data_structure}). This structure optimizes memory usage and access patterns, mitigating these challenges. It features two key parameters--$Capacity$ and $Size$--along with three supporting size-related parameters: $Size_{poly}$, $Size_{Modulus}$, and $Size_{Data}$. The $Size$ parameter, calculated as the product of these three supporting sizes, determines the memory allocation for a single encrypted data item.

$C\_Data$ instances dynamically allocate and manage memory to optimize performance. When data of $Size\_init$ arrives, the instance requests a corresponding memory block from the memory pool, setting both $Capacity$ and $Size$ to $Size\_init$. Upon expansion to $Size\_2$, the instance returns the initial memory and requests a new block of $Size\_2$, ensuring optimal memory usage. Conversely, when downsizing, the instance retains the existing memory to minimize data transfer overhead, adjusting $Size$ to the smaller value while $Capacity$ remains unchanged as $Size\_init$. This approach prioritizes time reduction over memory conservation.

\label{data_layout}

Within the memory, the data are logically stored in levels. Data bits belonging to the same modulus are stored continuously, one modulus after another. From a broader perspective, if a data number contains $n=Size_{Data}$ parts for longer bit storage, they will be stored separately for the convenience of calculation.

\subsubsection{Utilizing Multi stream}
\label{multi_stream}
CUDA streams offer a potent yet delicate mechanism for accelerating programs. They enable concurrent execution with explicit and carefully managed synchronization. To harness this capability responsibly, we've constructed a stream pool and meticulously integrated multi-stream operations into select operators. A common strategy involves assigning distinct streams to individual modulus, capitalizing on the contiguous data layout discussed in Section \ref{data_layout}. This modulus independence facilitates stream distribution and minimizes cross-modulus interference.

To prudently manage CUDA streams, the framework initializes a stream pool at the outset of each workload. The pool's size, denoted as $S_{stream}$, is determined by choosing the smaller one between the modulus chain length($modulus\_lenght$) and an empirically derived constant $MAX\_LEN$, typically set between 3 to 5. When computations can be executed independently on each modulus, the framework judiciously launches a maximum of $S_{stream}$ concurrent streams, ensuring optimal utilization of multi-stream capabilities. Subsequent synchronization is performed only when calculations hinge on the collective results of these streams. Section \ref{sec::eval} showcases the tangible performance gains achieved through this optimized stream management approach.

\subsection{Design of GPU-memory pool for \emph{CAT}}
\label{subsec-pool}

While exploiting GPUs for FHE computation unlocks substantial speed gains, it demands caution due to potentially ballooning memory requirements. This trade-off, made for ironclad data security, becomes particularly acute in large-scale tasks. For example, a single encrypted \emph{ciphertext}, with a default polynomial degree of 32768, occupies roughly 8 MB in its initial state. Unmanaged memory allocation in such scenarios can swiftly devour precious GPU DRAM, crippling performance and hindering large-scale FHE computations.

To address this issue, we have implemented a memory pool managing memory distribution and retrieval, allowing us to execute PDQ computing tasks composed of hundreds of operators using only 6 GB of GPU memory in serving mode.

\begin{figure}[htbp]
\centering
    \includegraphics[scale=0.35]{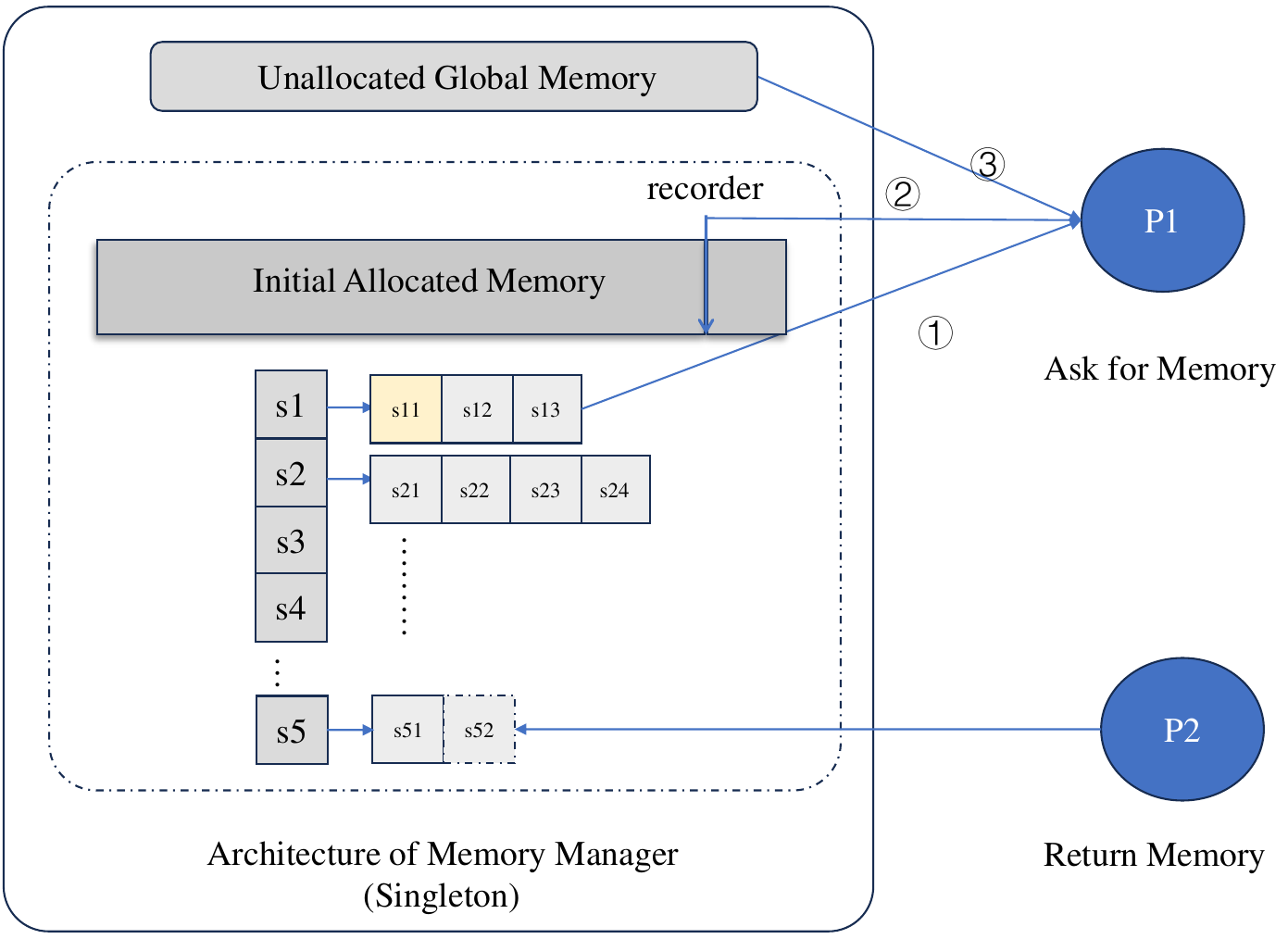}
\caption{Architecture of Memory pool}
\label{pool_architecture}
\end{figure}

\subsubsection{Architecture of Memory pool}
Let's begin by scrutinizing the architecture of the memory pool (refer to Fig. \ref{pool_architecture}). Our framework establishes an initial pool $P$ by allocating a contiguous memory segment with a size of $S$, calculated by Equation \ref{cal_memory_size}. This allocation is accomplished directly by invoking CUDA API. Subsequently, we establish a hash map to oversee the utilization of this allocated memory segment. Within the map, the key $s$ denotes the size of a slice of memory, and the corresponding value is a list of pointers indicating the starting points of continuous memory pieces with a size of $s$. Another pivotal element in the memory pool is the one-way \emph{recorder}, responsible for monitoring the extent of pool utilization. If we envision the initial memory as a linear railway, the \emph{recorder} progresses along the railway in a unidirectional manner and cannot backtrack.
\begin{equation}
\label{cal_memory_size}
 S = Min(2048, Length_{modulus}*200) \quad MB
\end{equation}

\subsubsection{Details about memory pool usage}
\label{memory_pool}

The memory pool primarily offers two essential functions: distributing and retrieving memory. We will elaborate on the processes involved in these two stages under three different conditions.

\textbf{Ask for memory in normal condition}
    
Upon receiving a request for memory of size $s_{need}$, the memory pool $P$ follows a systematic process. It begins by inspecting the value list associated with the key equal to $s_{need}$ in the hashmap. If the list is not empty, the last element $d$ in the list is allocated, along with the memory of size $s_{need}$ starting at position $d$. However, in scenarios where the list linked to the corresponding key is empty or the key is absent in the map, the position of the $recorder$ is designated, and the recorder, subsequently, advances forward for a length of $s_{need}$. This method ensures efficient allocation and utilization of memory resources based on the specified size requirements.

\textbf{Ask for memory when running out of the initial memory}

If the initial memory with size $S$ is depleted, new space for global GPU memory with size $s_{need}$ must be requested, rendering the recorder unnecessary from this point forward.

\textbf {Return memory}

In cases of memory shortage, when certain data $d$ completes its operation and will not be utilized in a subsequent operator, it is necessary to return the occupied memory $s_{return}$ to the memory pool. The return process is straightforward: $d$ only needs to link the starter of its memory to the value list of the key $s_{return}$. If the key does not exist, a new key and an empty list are generated to store it. It's noteworthy that the return process remains consistent, irrespective of the stage from which the request originates, as detailed earlier. This standardized approach ensures a systematic and efficient handling of memory resources throughout the framework.

\subsubsection{Discussions about the design of memory pool}

Several critical questions naturally arise regarding our memory pool design:

\textbf{Is memory pool really needed?}

\begin{figure}[htbp]
\centering
    \includegraphics[scale=0.145]{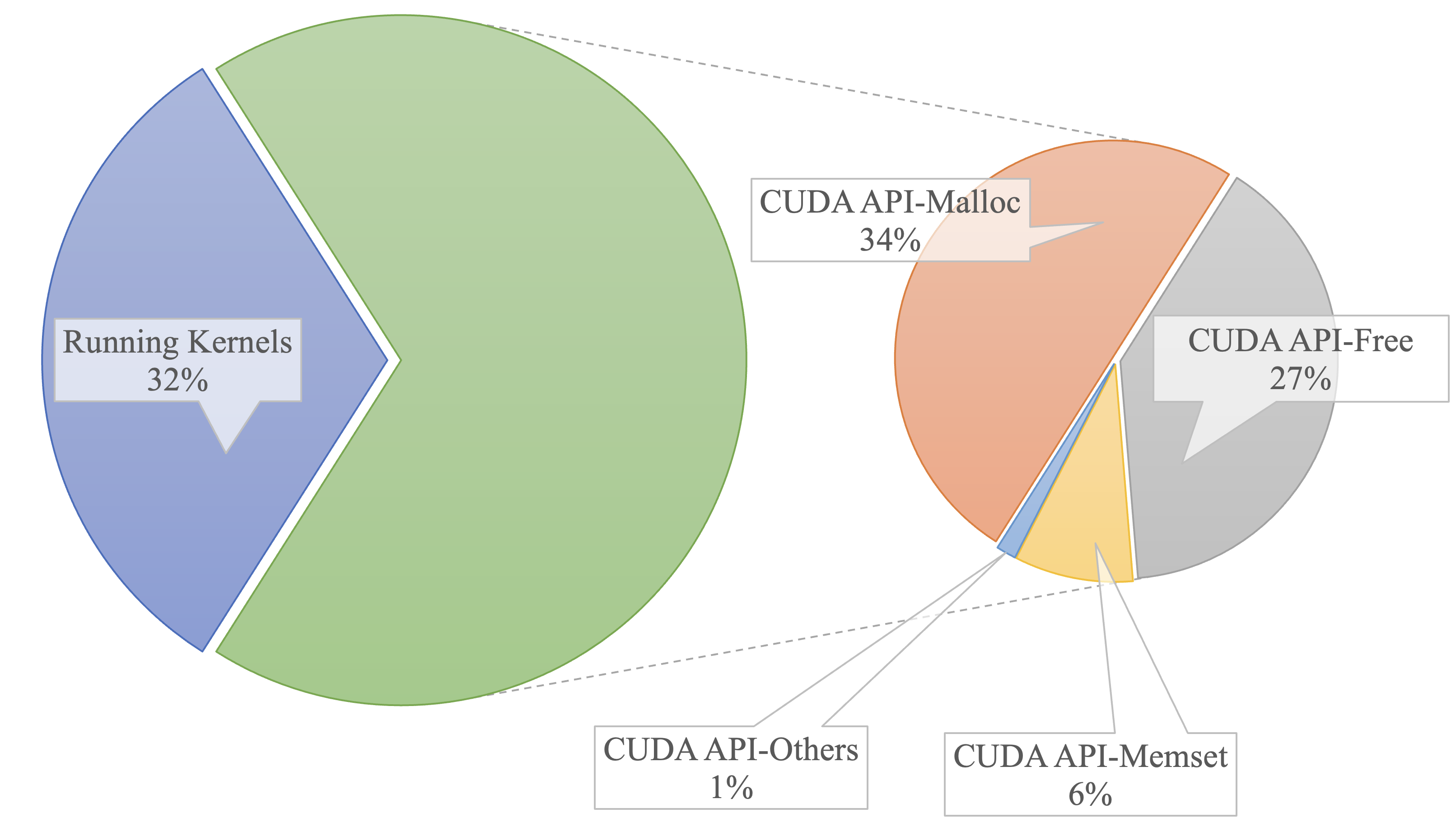}
\caption{Time consumption of each stages without any optimization}
\label{malloc_cost}
\end{figure}

A rudimentary solution to tackle memory shortages entails requesting and returning memory to the GPU using the CUDA API as needed. While this method is straightforward, it significantly hampers the overall computational efficiency, as illustrated in Fig \ref{malloc_cost}. The memory allocation and return stages alone account for more than 60\% of the time in the entire FHE program. Employing this method would essentially nullify the optimization efforts directed toward enhancing the computational operators. The adoption of a memory pool strategy becomes paramount in mitigating the computational overhead caused by frequent memory allocation and retrieval, 

\textbf{Why choose sizes as keys for the hash map?}

The CKKS/BFV/BGV schemes demonstrate that the size of encrypted and encoded data is determined by the multiplication of fixed parameters. This inherent property leads to the concentration of required memory within a limited range of size values. This trend extends beyond basic encrypted data to encompass intermediary parameters essential for operators. The structure of the memory pool can be likened to a retail dress shop exclusively stocked with fixed-size jackets. Customers simply select the size they need, and it fits seamlessly. In the realm of FHE, the necessity for a tailor to customize memory sizes for diverse needs becomes obsolete.

\textbf{Will the returned memory be wasted?}

\begin{figure}[htbp]
\centering
\subfigure[Return ratio]
{
    \begin{minipage}[b]{0.4\linewidth}
        \centering
        \includegraphics[scale=0.26]{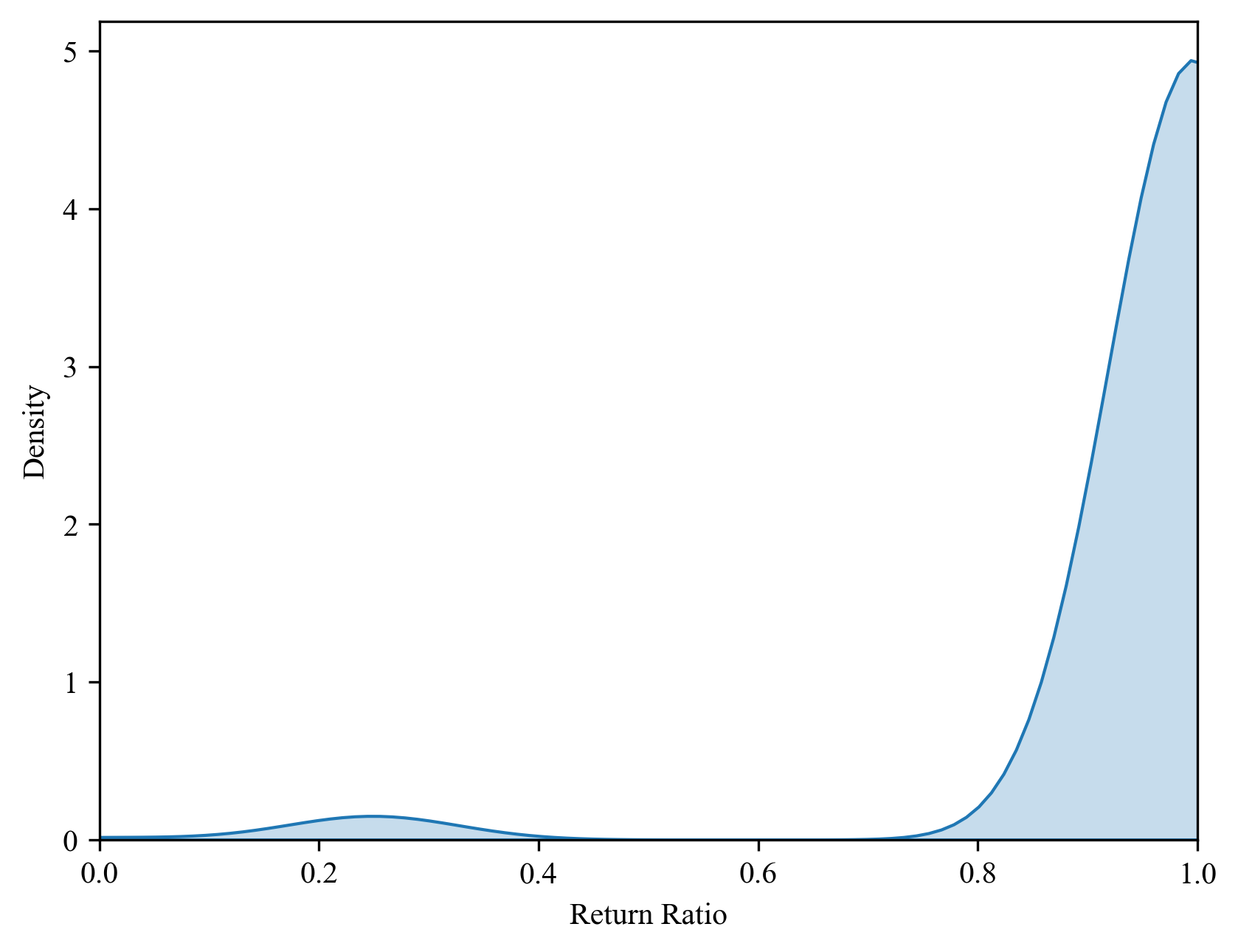}
    \end{minipage}
}
\subfigure[Reuse ratio]
{
 	\begin{minipage}[b]{0.4\linewidth}
        \centering
        \includegraphics[scale=0.26]{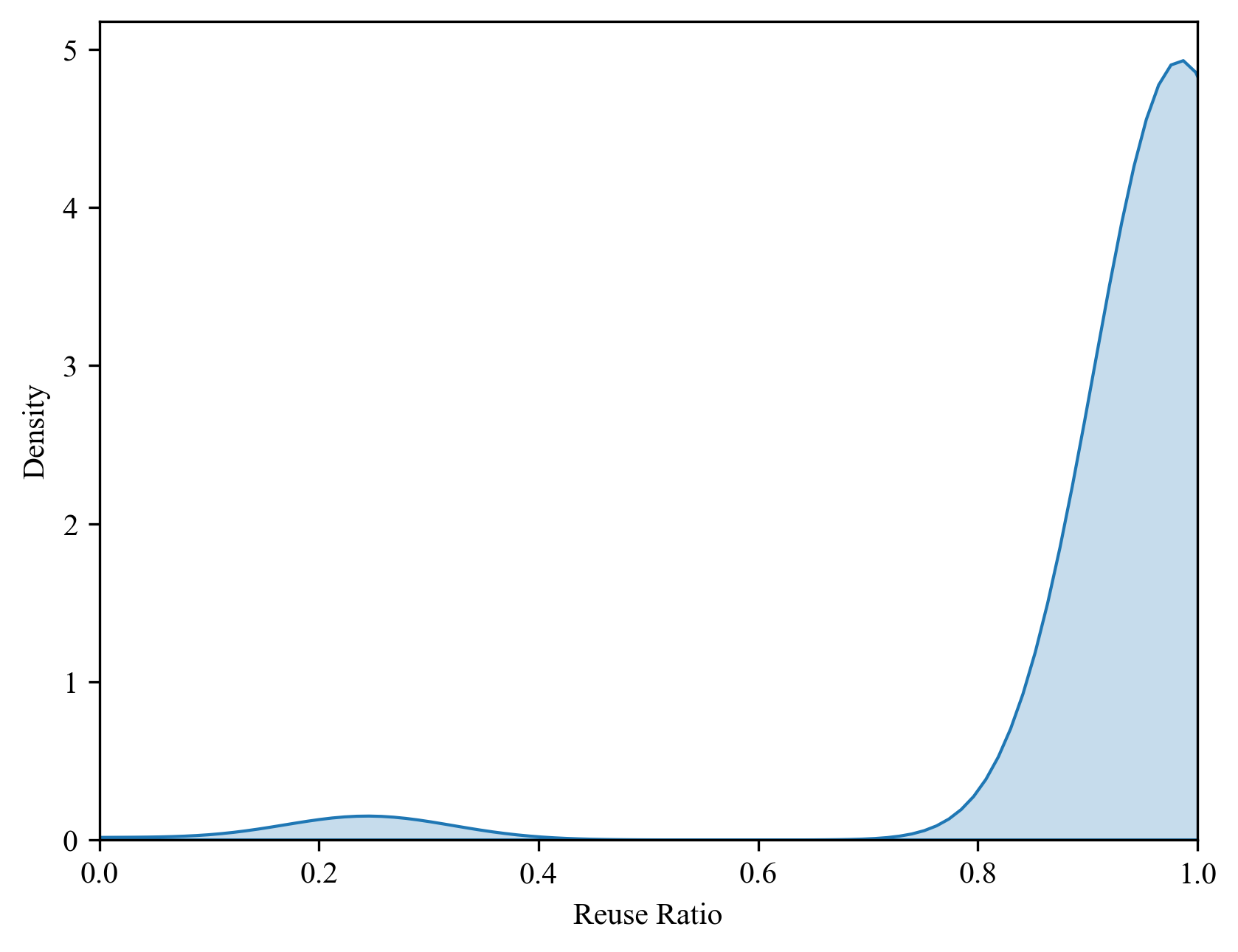}

    \end{minipage}
}
\caption{Status for memory return and reuse, pic a shows the density of memory blocks return ratio, calculated by return times divided by use times. pic b checks the times for memory askers get the memory needed from the pool}
\label{memory_use}
\end{figure}

Concerns may arise regarding potential imbalances in the hash map, particularly when certain keys result in much longer lists containing memory pieces used only once, because the mechanism we designed addresses this concern by ensuring that such memory remains unused until the entire program concludes. Our extensive experiments reveal a balanced distribution during both the ask and return stages for individual keys, as illustrated in Fig \ref{memory_use}. This equilibrium implies that memory asked more often will also be returned more frequently. Additionally, we've implemented a strategy to address potential imbalance. In cases where the map becomes unbalanced, during the asking stage, one can acquire a piece of memory larger than $s_{need}$ and closest to $s_{need}$ to mitigate this situation.

%% file: sec5-application.tex
We implemented the PDQ protocol based on CKKS, a scheme well-suited for multi-precision computations as discussed in~\cite{lee2023heaan}, to effectively validate our framework's capabilities in this scenario.
In this section, we first introduce our query execution framework in Section~\ref{subsec-pdqframework}. Then, we present the specific query types implemented in Section~\ref{subsec-instquery}. Finally, we give a detailed introduction of the search-and-average query based on our newly proposed two-party division protocol in Section~\ref{subsec-searchsumquery}.

\subsection{Privacy Database Query Framework}\label{subsec-pdqframework}
The PDQ protocol aims to provide the client with a set of
records that match its query while maintaining security. 
We adopt the same approach as in~\cite{10376195} for the security model and threat model, which guarantees both query privacy and database privacy under the semi-honest threat model. 
This model assumes that both the server and the user adhere to the protocol specifications, but they may attempt to learn additional information from the exchanged data.

\begin{figure}[!h]
\centering
    \includegraphics[width=8.3cm]{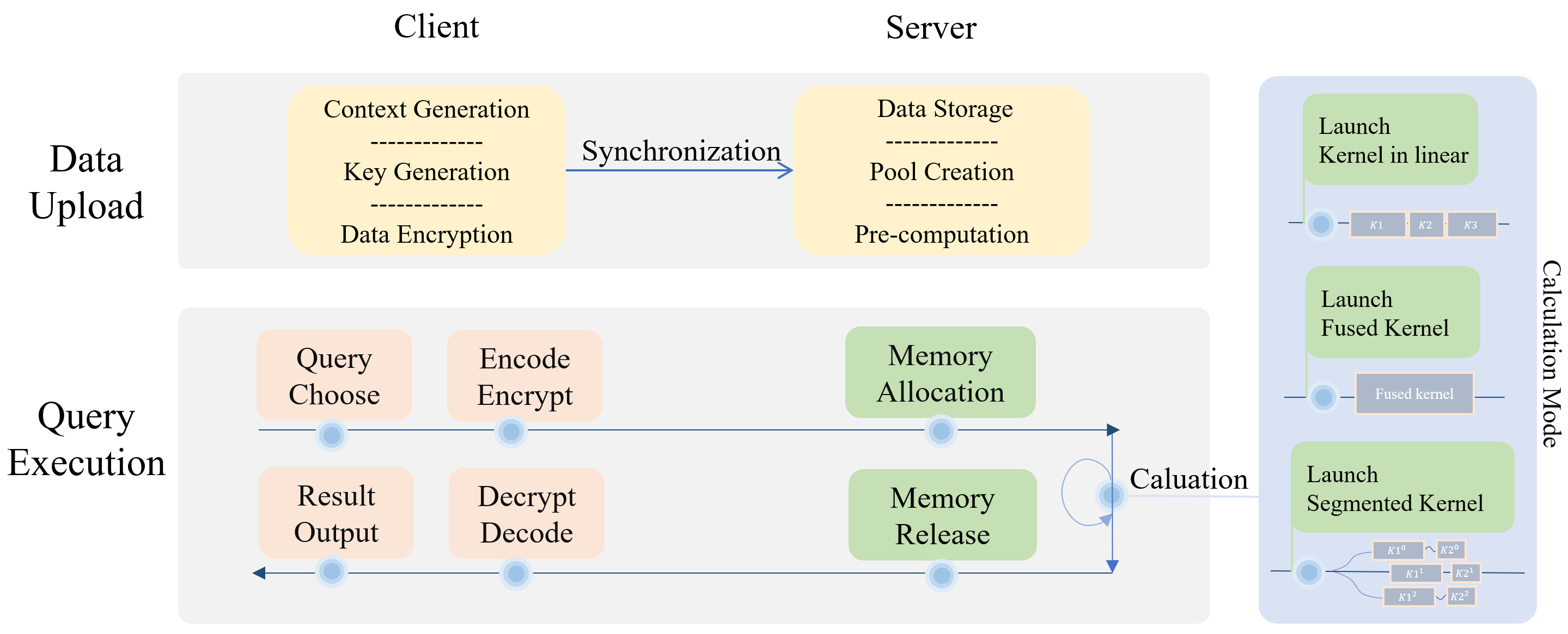}
\caption{Query Execution Process}
\label{query_process}
\end{figure}

The entire framework execution is divided into two stages: Data Upload and Query Execution, as depicted in Figure~\ref{query_process}.

In the first stage, the client generates the context that meets all query requirements. 
This includes selecting the appropriate RLWE parameters $q$ and $n$, generating the keys, encrypting the data and uploading the ciphertext to the server, and synchronizing the context and public parameters with the server.
After that, the server generates the memory pool and stream pool for all needed foundational operations based on the information uploaded by the client, as well as other parameters that can be pre-computed, such as the NTT table.

In the second stage, for each query process, the client encodes and encrypts the specific query condition and sends it to the server. 
The server stores the encrypted query in GPU memory allocated from the memory pool, and remains there for the query's duration. 
The framework optimizes calculation modes (linear, fused, segmented, or combinations) for efficient and accurate computations. 

Upon completion, encrypted results will be transfered to CPU memory are sent back to the client, while preserving all the context including pre-computed parameters and the dual GPU pools. The client then gets the final result by decrypting and decoding the received ciphertext.

\subsection{Instantiated Queries}\label{subsec-instquery}
Our implementation of the query function follows the methodology proposed in~\cite{10376195}. 
The search part is implemented by first decomposing the clear text by a small base $p$ and then performing an FFT transformation to complete the pre-processing before encoding. 
The resulting values are then used to execute the comparison with specific encrypted query condition values using the Lagrange Polynomial Interpolation technique. 
The computation part is implemented by performing subsequent linear and non-linear operations on the results of the search part.

In~\cite{10376195}, they implement three queries: 
1. The search query implementation extracts the data that meets specific comparison-like conditions, by multiplying the comparison results ($1$ or $0$) with the data; 2. The search-and-sum query can be seen as performing an addition operation on the result of the search query; 3. The BMI query, which in clear text is $height/weight\cdot weight$, performs a square and division operation on the data that meets specific conditions.

We implement all these queries and give a detailed performance comparison with~\cite{10376195} in Section~\ref{sec::eval}. 

\subsection{Search-and-Average Query}\label{subsec-searchsumquery}

Motivated by that the bootstrapping-based multiply-inverse ($1/x$) operation in~\cite{lee2023heaan} is inefficient, and the method adopted in~\cite{10376195,DBLP:conf/asiacrypt/CheonKKS17} requires the range of $x$ to be within $p/2$ to $3p/2$ for some known $p$ to ensure accuracy, we present a new multiply-inverse method based on multiplicative masking.

Our new method assumes that the constraint\footnote{In ~\cite{lee2023heaan}, the author's constraints that \emph{after
sharing data, they must not participate in the analysis
process.}} in ~\cite{lee2023heaan}, which is not strictly necessary for our targeted single-data-holder outsourcing scenario. Therefore, the encrypted intermediate results can be returned to the data holder, while ensuring that the data holder does not perform any ciphertext evaluation operations, thus complying with the requirements of FHE-based PDQ.
The two-party computation process is as shown in Algorithm~\ref{alg-mulinverse}.

\begin{algorithm}[!h]
\caption{The Two-Party Multiply-Inverse Method}
\label{alg-mulinverse}
\begin{algorithmic}[1] % The number tells where the line numbering should start
\Procedure{Multiply-Inverse}{}
    \State \textbf{Input: $C_0 = Enc(x)$} 
    \State \textbf{Output: $C_2 = Enc(\frac{1}{x})$} 
    \Statex
    \State \textbf{Server:}
    \State \textit{Choose random number $r$ and compute $C_1 = C_0\cdot r$}
    \State \textit{Send $C_1$ to Client}
    %\State \indent{Choose random number $r$ and compute $C_1 = C_0\cdot r$ Statement(s) for part 1 executed by Party A}
    \Statex
    \State \textbf{Client:}
    \State \textit{Decrypt and get $x_1 = Dec(C_1)$}
    \State \textit{Compute $x_2 = \frac{1}{m_1}$}
    \State \textit{Encrypt $x_2$ and send $C_1 = Enc(x_2)$ to Server}
    
    %\State \indent{Statement(s) for part 2 executed by Party B}
    \Statex
    \State \textbf{Server:}
    \State \textit{Compute $C_2 = C_1\cdot r = Enc(\frac{1}{x})$}
    %\State \indent{Statement(s) for part 3 executed by Party A}
\EndProcedure
\end{algorithmic}
\end{algorithm}

\textbf{Privacy Security:} 
The Server is only able to learn the ciphertext values, while the Client is only able to learn the plaintext value of $r\cdot x$, which is masked by a multiplication mask. Therefore, the plaintext value of $x$ or $1/x$ is not leaked to either Client or Server.

Our method can hold significance when the cost of each interaction is substantially lower than that of performing a bootstrapping operation.
And based on this new multiply-inverse method, we achieve a new query function: search-and-average.
The experiment result shows that all query processes can be completed within one second. We will elaborate more in Section~\ref{sec::eval}.

%\begin{figure}[!h]
%    \centering
%    \includegraphics[width = 0.45\textwidth]{drawings/new-choice.png}
%    \caption{Choices of High-Depth Operation Process}
%    \label{fig:new-choice}
%\end{figure}

%% file: sec6-evaluation.tex
\begin{table*}[t]
\caption{Running time in $\mu$s and speed up times for all CKKS operators}
\label{ckks-operator}
    \begin{tabular*}{\linewidth}{@{\extracolsep{\fill}}llrrrrr}
    \toprule
                                       & Operator                  & SEAL(CPU) & \emph{CAT}(3090) & \textbf{SpeedUp} & \emph{CAT}(4090) & \textbf{SpeedUp} \\ \hline
     
                                       & encode                    & 12387     & 2477           & \textbf{5x}    & 1083            & \textbf{11x}    \\
                                       & decode                    & 59041     & 3691           & \textbf{16x}    & 1718            & \textbf{34x}    \\
                                       & encrypt                   & 71886     & 1432            & \textbf{50x}   & 1294            & \textbf{56x}    \\
     \multirow{6}{*}{ CKKS}
                                       & decrypt                   & 4472      & 27              & \textbf{166x}  & 20             & \textbf{223x}    \\
   \multirow{7}{*}{ N=32768, L=16, Q=881}                                    

                                       & add                       & 3954      & 7               & \textbf{565x}  & 
                                       2              & \textbf{1977x}    \\
                                       & multiply                  & 8187      & 32              & \textbf{256x}  & 23              & \textbf{356x}   \\
                                       & multiply\_plain           & 3374      & 4               & \textbf{844x}  & 2               & \textbf{1687x}  \\
                                       & square                    & 8269      & 26              & \textbf{318x}  & 17              & \textbf{486x}   \\
                                       & relinearize               & 175520    & 3607            & \textbf{49x}   & 2617            & \textbf{67x}    \\
                                       & rescale                   & 20132     & 687             & \textbf{29x}   & 440             & \textbf{46x}    \\
                                       & rotate\_vector\_one\_step & 175741    & 3595            & \textbf{49x}   & 2577            & \textbf{68x}    \\
                                       & rotate\_vector\_random    & 876642    & 18440           & \textbf{48x}   & 11837           & \textbf{74x}    \\
                                       & complex\_conjugate        & 174508    & 3495            & \textbf{50x}   & 2556            & \textbf{68x}    \\ 
                                       \bottomrule 
    
  \end{tabular*}
\end{table*}

\begin{table*}[t]
\caption{Running time in $\mu$s and speed up times for all BFV operators}
\label{BFV-operator}
\begin{tabular*}{\linewidth}{@{\extracolsep{\fill}}llrrrrr}
    \toprule
                                        & Operator                  & SEAL(CPU) & \emph{CAT}(3090) & \textbf{SpeedUp} & \emph{CAT}(4090) & \textbf{SpeedUp} \\ \midrule   
                                        & encode\_batch             & 680       & 366             & \textbf{2x}     & 339             & \textbf{2x}     \\
                                        & encrypt                   & 57018     & 696         & \textbf{82x}     & 427           & \textbf{136x}     \\
        \multirow{6}{*}{ BFV}
                                        & decrypt                   & 25688     & 662          & \textbf{39x}     & 471            & \textbf{55x}     \\
        \multirow{7}{*}{ N=32768, L=16, Q=881}
                                        & add                       & 4130      & 12              & \textbf{344x}   & 3              & \textbf{1378x}    \\
                                        & multiply                  & 306318    & 4280            & \textbf{72x}    & 3675            & \textbf{83x}    \\
                                        & multiply\_plain           & 42808     & 762             & \textbf{56x}    & 556             & \textbf{77x}    \\
                                        & square                    & 230590    & 7470            & \textbf{31x}    & 4157            & \textbf{55x}    \\
                                        & relinearize               & 181471    & 7800            & \textbf{23x}    & 4607            & \textbf{39x}    \\
                                        & rotate\_vector\_one\_step & 181883    & 3643            & \textbf{50x}    & 2633            & \textbf{69x}    \\
                                        & rotate\_vector\_random    & 808661    & 15882           & \textbf{51x}    & 11299           & \textbf{72x}    \\
                                        & rotate\_columns           & 181205    & 3275            & \textbf{55x}    & 1623            & \textbf{112x}   \\ \bottomrule
    
    \end{tabular*}
\end{table*}

\begin{table*}[!h]
\caption{Running time in $\mu$s and speed up times for all BGV operators}\label{bgv-operator}
% \begin{tabular*}{\textwidth}{@{\extracolsep{\fill}}llrrrrr}
\begin{tabular*}{\linewidth}{@{\extracolsep{\fill}}llrrrrr}
\toprule
                                   & Operator                  & SEAL(CPU) & \emph{CAT}(3090) & \textbf{Speed} & \emph{CAT}(4090) & \textbf{Speed} \\ \hline

                                   & encode\_batch             & 640     & 466           & \textbf{1x}    & 371            & \textbf{2x}    \\
                                   & encode\_unbatch           & 773     & 527           & \textbf{1x}    & 480            & \textbf{2x}    \\
                                   & encrypt                   & 98133     & 1432            & \textbf{50x}   & 1294            & \textbf{76x}    \\
   \multirow{6}{*}{ BGV} 
                                   & decrypt                   & 19376      & 762              & \textbf{25x}  & 519             & \textbf{37}    \\
                                   
   \multirow{7}{*}{ N=32768, L=16, Q=881}
                                   & add                       & 4053      & 10               & \textbf{405x}  & 6              & \textbf{676x}    \\
                                   & multiply                  & 8762      & 23              & \textbf{381x}  & 17              & \textbf{515x}   \\
                                   & multiply\_plain           & 15433      & 420               & \textbf{37x}  & 191               & \textbf{81x}  \\
                                   & square                    & 10865      & 6              & \textbf{1810x}  & 5              & \textbf{2173x}   \\
                                   & relinearize               & 193632    & 4979            & \textbf{39x}   & 3169            & \textbf{61x}    \\
                                   & rotate\_rows\_one\_step & 203200    & 4182            & \textbf{49x}   & 2981            & \textbf{68x}    \\
                                   & rotate\_rows\_random    & 872223    & 16847           & \textbf{52x}   & 11837           & \textbf{74x}    \\
                                   & rotate\_columns    & 196284    & 4018           & \textbf{49x}   & 2786           & \textbf{70x}    \\
                                   \bottomrule 

\end{tabular*}
% \end{sidewaystable}
\end{table*}

We employ a series of benchmarks to demonstrate the efficacy of our framework. Including a designed script that contains almost all the basic operators, and a few functions in the scenario of PDQ to demonstrate realistic computational workloads. We run most of the experiments on an Nvidia 4090 GPU and an AMD EPYC 9654 CPU by default to show the best speed-up that we can achieve. Moreover, we have run experiments on different GPUs like Nvidia 3090, Nvidia V100, etc. to reveal the relationship between the running speed and the hardware capacity. 

\subsection{Basic FHE operators}
Basic FHE operators mainly contain 5 stages: Encode, Encrypt, Calculation, Decrypt, Decode, along with Creating several keys. We have built a performance-testing script that contains all the operators that an FHE program may use. We set the degree of poly(N) to 32768, the length of modulus chain(L) to 16, and the total bit count(Q) to 881 by default and test the performance for CKKS, BFV, and BGV scheme respectively \footnote{The implementation of BGV scheme is similar to CKKS's and so is the test result, we attach the result table in the appendix.}. We ran each test 10 times and recorded the average time each operator used. The results are shown in Table \ref{ckks-operator}, Table \ref{BFV-operator} and Table \ref{bgv-operator}. As can be seen from the results, our framework can achieve up to 2173$\times$ when comparing the running time of common operators in FHE with Microsoft's SEAL. We achieve at least 10 times acceleration throughout all stages including encoding, encrypting, and computation, which makes the PDQ application built on FHE 20 times faster.

Besides, we have compared our framework with some early works focused on the acceleration of FHE with GPU. Most of the source codes are not publicly available, we can only compare a few of open sourced operators (Hmult, Cmult, Add, Rescale, Rotate) by \cite{jung2021over} with our framework under default configuration, and then using the relative comparison result given by other works\cite{DBLP:conf/hpca/FanWXHMZ23,al2020privft} to estimate their performance.\footnote{In terms of achieving a balance between efficiency and capability in PDQ, we restrict the modulus length in \emph{CAT} under 17. While \cite{DBLP:conf/hpca/FanWXHMZ23,al2020privft} only release their relative strength over \cite{jung2021over} under 44, thereby rendering our comparison a reasonable estimation..}. We summarize the result in Table \ref{compare_other_gpu}, and the results shows in the table is how many times computation time will be consumed individually if we set the time cost of our framework as $1$. It can be seen that we outperform other works with large gaps. The result shows that our framework has a promising acceleration ability.

\begin{table}[]
\caption{Compare with other implementations on several operators}
\label{compare_other_gpu}
% \tabcolsep=0.018\linewidth
\begin{tabular*}{\linewidth}{@{\extracolsep{\fill}}lrrrrr}
\toprule
          & \multicolumn{1}{l}{Hmult} & \multicolumn{1}{l}{Cmult} & \multicolumn{1}{l}{Add} & Rescale & Rotate  \\ \midrule

TensorFHE & 1.33                               & 1.13                             & 1.29          &  1.85 & 1.27                  
\\
% 100x      & 64                               & 17                               & 20.5                    & 2.62                       & 7.12                        \\
100x      & 1.39                               & 3.25                               & 5.54    &19.5 &1.30                                    \\

% PrivFT    & 450                              & 9.18                             & 41                      & /                          & 9.69                        \\
 PrivFT    & 4.8                              & 1.755                            & 2.77      & 42.9 & -                                    \\
\emph{CAT}      & \textbf{1}                       & \textbf{1}                       & \textbf{1}       & \textbf{1}                       & \textbf{1}                                  \\ \bottomrule
\end{tabular*}
\end{table}

\subsection{CKKS-based PDQ Application}

Benefiting from the GPU acceleration of our implementation and the newly proposed multiply-inverse method, our PDQ implementation addresses the challenges described in Section~\ref{subsec-pdq} and fulfills the requirements of functional completeness, accuracy, and efficiency, without employing bootstrapping or scheme switching. 
Our query can support data range up to $2^{32}$, where the data can be integers or real numbers. 
We ensure integer part accuracy for all query results, meaning multiply-inverse operations $\frac{1}{x}$ have a 32-bit binary accuracy beyond the decimal point with no range limitation of $x$.

% \textcolor{red}{
As introduced in Section \ref{sec::app-PDQ}, we test our framework on four different types of PDQ queries, recording its running time and memory usage shown in Table \ref{end-to-end result}. Since there is no open-source industrial ready GPU framework for FHE to the best of our knowledge, we can only compare the result with the implementation using Microsoft SEAL on CPU. Besides, we have also compared with \cite{10376195} as mentioned in \ref{subsec-instquery} on PDQ with CPU on certain cases which \cite{10376195} have provided their results. The query we evaluated are as listed and the number of data rows is 1024, the $Col_x$ represents the $x$ column in the dataset, Sum() and Avg() are the operations for summation and average, and the Index() simply fetches the ids of the rows that meet the filters.
% }
\begin{itemize}
    \item Index() where ($Col_a <= Col_b$ and $Col_c != Col_d$)
    \item Sum($Col_a$) where ($Col_b <= Col_c$ and $Col_d != Col_e$) 
    \item $Col_a$ / $Col_b^2$ where ($Col_b <= Col_c$)
    \item Avg($Col_a$) where ($Col_b <= Col_c$ and $Col_d == Col_e$) 

\end{itemize}

% \textcolor{red}{
Table \ref{end-to-end result} presents the results, the "Cal" in the first two columns indicates the pure calculation stage, and "Full-Task" in the third column consists of the time for Encryption, Decryption, Calculation, and Serialization to simulate the query application in real world. "1st Cal" shows the time needed for running the first time, it would be slower than later ones because of the process of building the memory pool and segmentation memory into pieces. It can be seen from the Table that our framework achieves up to 33x acceleration compared to the implementation on CPU, and the time needed for the entire task is no more than \textcolor{red}{1} second, making it possible to be deploymented into real-world applications. Moreover, Table \ref{end-to-end memory} records the memory occupation on GPU when running each PDQ task, as shown in the result that the queries process about 2$ \sim $ 5 GB to finish the entire task. 
% }

\begin{table}[h]
\caption{Running time in $m$s and speed up times for running PDQ tasks}
\label{end-to-end result}
\begin{threeparttable}          
\begin{tabular*}{\linewidth}{@{\extracolsep{\fill}}clrrr}
\toprule
\multicolumn{1}{l}{}                       &         & 1st Cal                    & Cal                     & Full-Task\tnote{1}                    \\ \midrule
\multirow{4}{*}{PDQ-1}                     & CPU     & 3173                              & 2971                   & 3387                           \\
                                           & \cite{10376195}     & 5200\tnote{2}                              & 5200                   & 52702\tnote{3}                         \\
                                           & \emph{CAT}     & 179                               & 167                     & 241                            \\
                                           & SpeedUp & \textbf{18x}                      & \textbf{18x}            & \textbf{14x}                   \\\midrule
\multirow{4}{*}{PDQ-2}                     & CPU     & 9430                              & 9198                    & 9715                           \\
                                           & \cite{10376195}     & 3220\tnote{2}         & 3220           & 3500\tnote{3}                   \\
                                           & \emph{CAT}     & 533                               & 357                     & 395                            \\
                                           & SpeedUp & \textbf{18x}                      & \textbf{26x}            & \textbf{25x}                   \\ \midrule
\multirow{4}{*}{PDQ-3}                     & CPU     & 8.7s                            & 8.4s                    & 9.1s                           \\
                                           & \cite{10376195}     & 14.58s\tnote{2}                            & 14.58s                    & 15s\tnote{3}                           \\
                                           & \emph{CAT}     & 291                               & 254                     & 377                            \\
                                           & SpeedUp & \textbf{30x}                      & \textbf{33x}            & \textbf{24x}                   \\  \midrule
\multirow{3}{*}{PDQ-4}                     & CPU     & 16.8s                              & 16.1s                   & 17.4s                           \\
                                           & \emph{CAT}     & 977                               & 887                     & 2.4s                            \\
                                           & SpeedUp & \textbf{18x}                      & \textbf{18x}            & \textbf{7x}                   \\

                                           \bottomrule
\end{tabular*}
\begin{tablenotes}    %这行要添加， 从这开始
        \footnotesize               %这行要添加
        \item[1] The Full-Task does not take the time of transferring data through into consideration.
        \item[2] The 1st Cal for \cite{10376195} is set equal to Cal.          %这行要添加
        \item[3] The Full-Task of \cite{10376195} is estimated.         %这行要添加
      \end{tablenotes}            %这行要添加

\end{threeparttable}
\end{table}

\begin{table}[h]
\caption{GPU Memory Cost for running PDQ task in MB}
\label{end-to-end memory}
% \tabcolsep=0.03\linewidth
\begin{tabular*}{\linewidth}{@{\extracolsep{\fill}}lrrrr}
\toprule
                & PDQ-1 & PDQ-2 & PDQ-3 & PDQ-4 \\ \midrule
GPU Mem Cost & 2316     & 2521     & 2191  & 5429     \\ \bottomrule
\end{tabular*}
\end{table}

\subsection{Ablation Analysis}

Since we have tested the framework we proposed for its efficiency on the operator level and high-precision PDQ application as a whole, we would like to dig deeper into how the key variants would influence the running performance.

\begin{table}[h]
\caption{Hardware Parameters of Different GPUs}
\label{gpu-params}
% \tabcolsep=0.038\linewidth
\begin{tabular*}{\linewidth}{@{\extracolsep{\fill}}lllll}
\toprule
\multicolumn{1}{c}{}  & \multicolumn{1}{c}{}   & \multicolumn{2}{c}{\begin{tabular}[c]{@{}c@{}}Compuation \\ Power(TFLOPS)\end{tabular}} & \multicolumn{1}{c}{}                               \\ \cmidrule(lr){3-4}
\multicolumn{1}{c}{\multirow{-2}{*}{Type}} & \multicolumn{1}{c}{\multirow{-2}{*}{\begin{tabular}[c]{@{}c@{}}Memory\\ (GB)\end{tabular}}} & Single                                                                & Half      & \multicolumn{1}{c}{\multirow{-2}{*}{Architecture}} \\ \midrule
1080Ti  & 11                                           & 11.34 & 11.34 & Pascal                         \\
3090Ti  & 24                                           & 35.58                                                                           & 71    & Ampere                         \\

A100                   & \textbf{40}                                           & 19.5                                                                            & \textbf{312}   & Ampere                         \\
V100                   & 26                                           & 15.7                                                                            & 125   & Volta                          \\ 
4090                   & 24                                           & \textbf{82.58}                                                                           & 165.2 & Ada Lovelace                   \\
\bottomrule
\end{tabular*}
\end{table}

\begin{figure}[h]
\centering
    \includegraphics[scale=0.33]{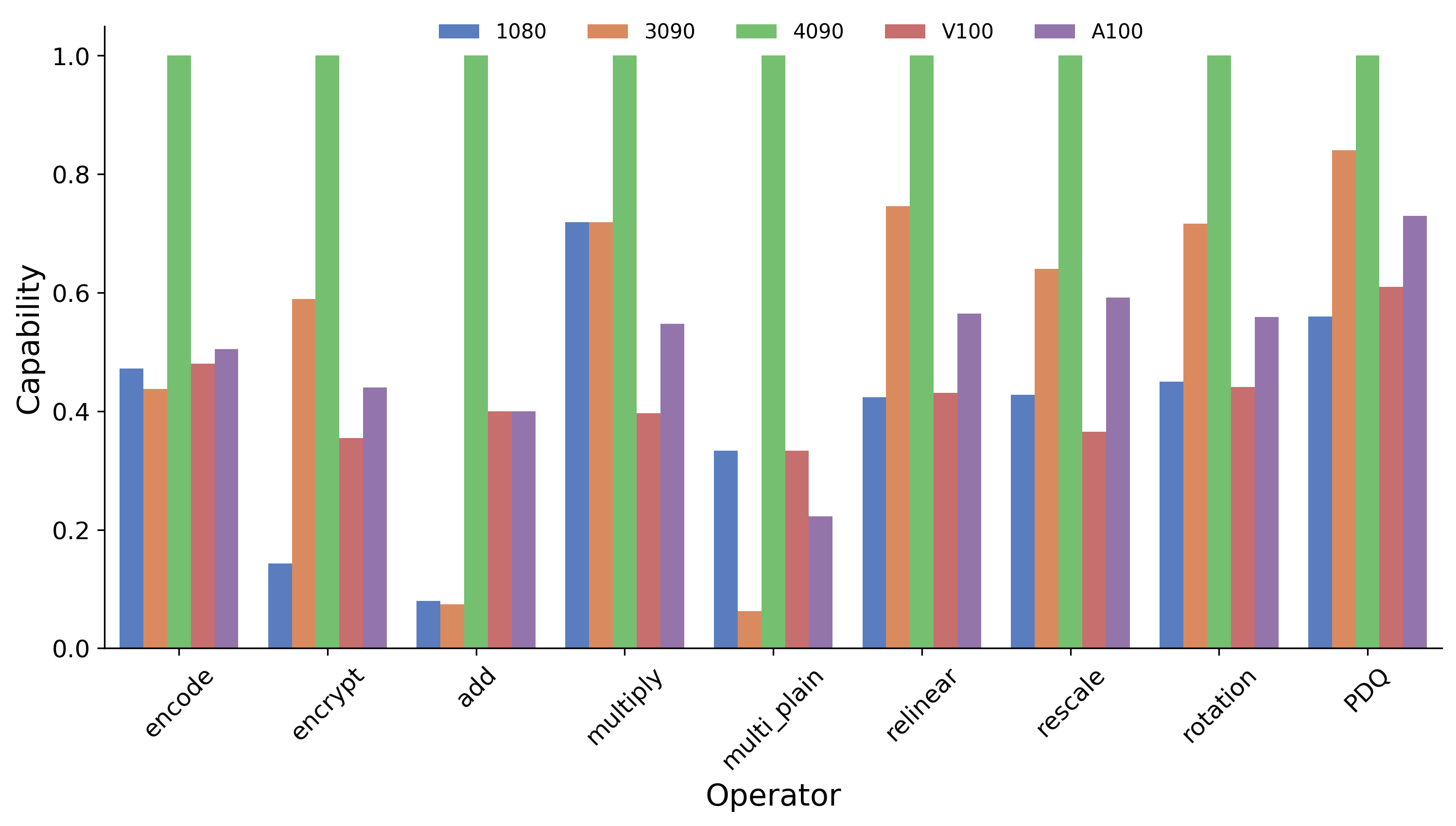}
        % \label{ablation:memory}
\caption{Computation capability of different type of GPUs}
\label{ablation-diff-gpu}
\end{figure}

\begin{figure}[h]
\centering
    \includegraphics[scale=0.365]{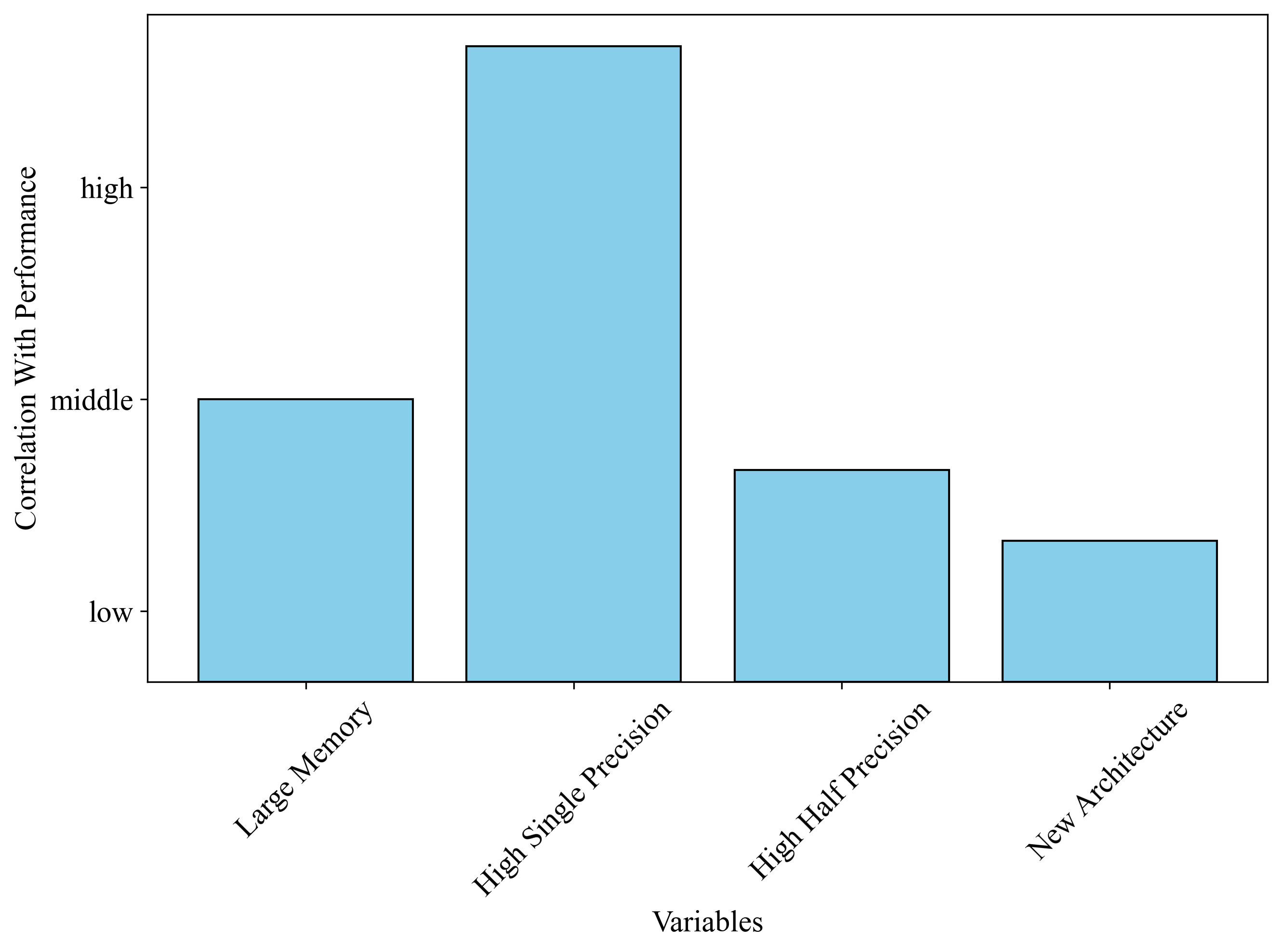}
        % \label{ablation:memory}
\caption{Correlation between computation capability on FHE with GPUs' hardware Variables}
\label{ablation-corr-hard}
\end{figure}

\textbf{Different GPUs}

The GPU card one uses would intuitively affect the running time consumed. We have tested all the operators and the PDQ-1 workload on 5 different types of GPU cards, including 4 major hardware architectures shown in Table \ref{gpu-params}. As shown in Fig \ref{ablation-diff-gpu}, the program runs fastest on Nvidia 4090, which reaches the highest single precision computation capacity. We use the $\frac{1}{t_{compute}}$ to represent the power of performing FHE computation, the higher the value, the more capable the card is. We can notice that under most situations, none of the other types of card can outperform 4090 because of its overwhelming superiority of computation ability on single precision. And it shows a high correlation between single precision computation power and its capacity on FHE. We can draw some other conclusions shown in Fig \ref{ablation-corr-hard} that the computation power of half-precision and architecture types have relatively less relationship with the computation ability of FHE. While the memory has some impact because of that the threshold of a certain amount of memory is needed for running the whole workload, once the threshold has been met, the excess memory does not bring more advantages for running FHE workloads with our framework.

It is delightful to observe that even a budget card like the 1080Ti can finish all the tests including the PDQ task, and even reach about 60\% of the fastest Nvidia 4090 in the PDQ scenario. It shows the great generality of our framework and makes acceleration for industrial FHE applications more reachable just with a card having obsolete architecture and computation power.

% \textbf{Kernel running mode}

% We have set three logical modes for kernels running at different part in our framework. We would like to test how much improvement have been brought by running the PDQ application with all kernels arranged linearly and a combination of three modes. The results show that a set of several appropriate modes brings x\% of alleviation.

\textbf{Memory pool}

\begin{figure}[htbp]
\centering
    \includegraphics[scale=0.3]{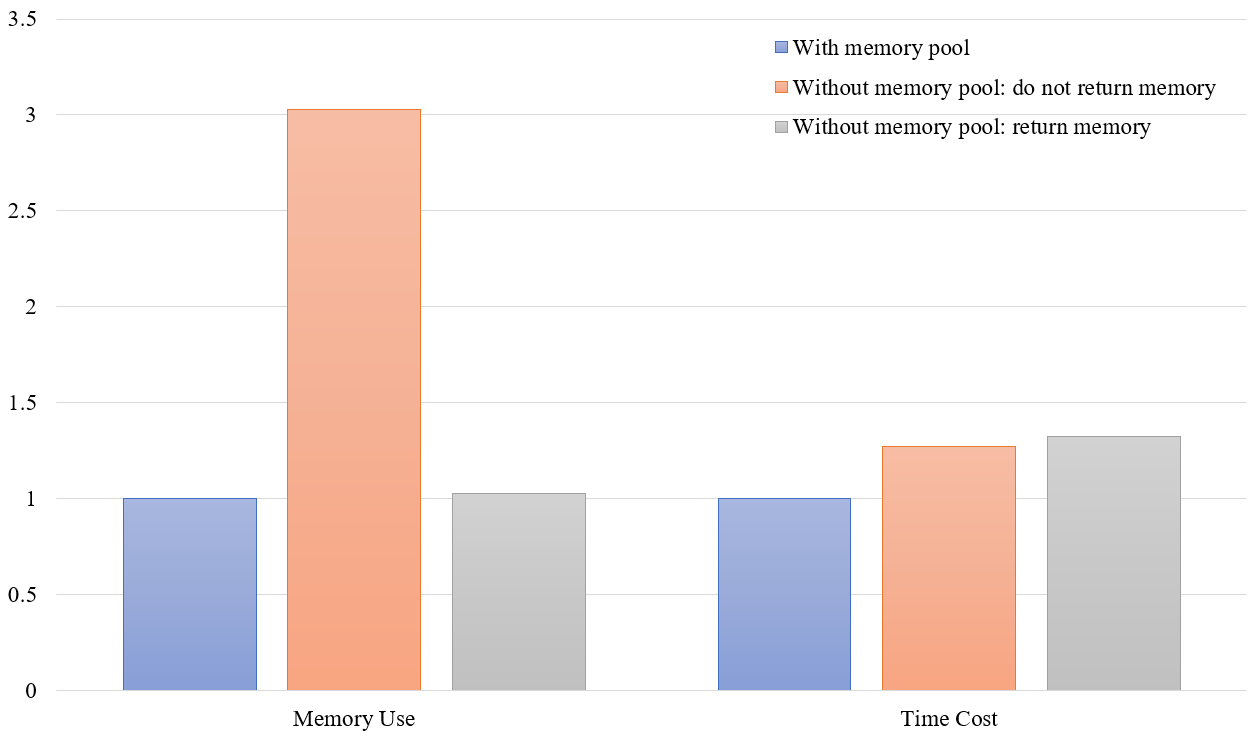}
        % \label{ablation:memory}
\caption{Memory usage and time cost under different running mode}
\label{ablation-memorypool}
\end{figure}

Bringing and developing the GPU memory pool is the key to our framework can run the workload costing less memory and maintain the speed of calculating. There are two ways to manage the memory without the pool, by ignoring the returning stage and returning it back to the device every time. The first way will not add any additional time to the running workload, but the speed of consuming memory is significantly fast, while the second way will not waste any part of memory, but the extra time cost for returning the memory will slow the whole workload down. The numerical result in Fig \ref{ablation-memorypool} shows that the first way will consume 3 times more memory to finish the whole computation, and the second way costs 30\% more time for the entire task.

\textbf{Stream Pool}

\begin{figure}[h]
\centering
    \includegraphics[scale=0.2]{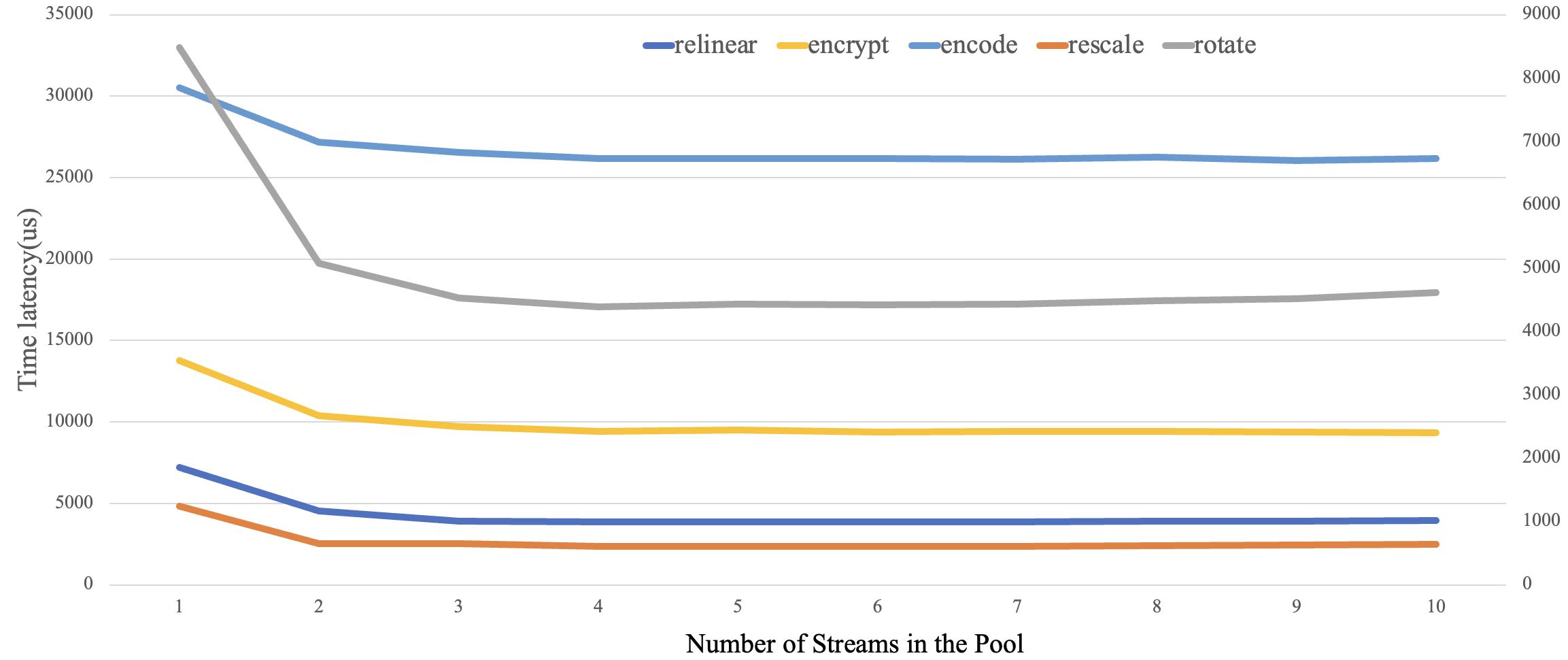}
\caption{Time latency of running operators without and with stream pool containing different number of streams}
\label{ablation-streampool}
\end{figure}

We have theoretically shown the advantage of using multiple streams brought by overlapping several independent workloads while consuming more hardware resources at the same time. We test the time latency by running several complicated operators with and without stream pool(using only one stream), additional with different stream numbers. The result in Fig \ref{ablation-streampool} shows the introducing of the stream pool will reduce at most 40\% latency to the whole computation, and a pool with more than 3 streams will achieve the best result.

\textbf{Mod for Big Integers}

\begin{figure}[h]
\centering
    \includegraphics[scale=0.275]{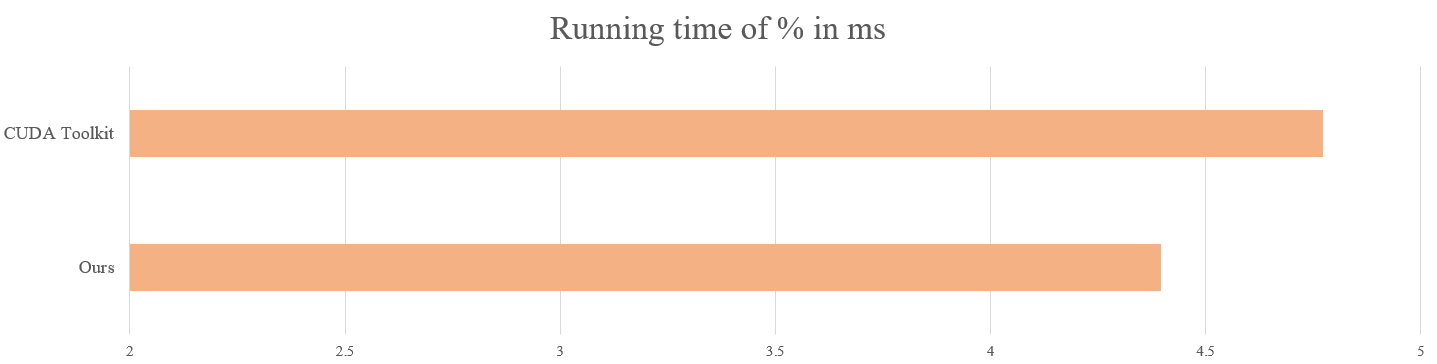}
\caption{Running time in ms for $\%$ with our implementation and CUDA toolkit's}
\label{ablation-percent}
\end{figure}

The implementation of $\%$ for big integers we presented in Algorithm \ref{module_gpu} contributes to the workflow. We run a comparison between our implementation and the embedded operator $\%$ in CUDA for big integers. We run $10^8$ times $X \% Y$ with two implementations individually, and the numerator and denominator are both set larger than 32-bit and smaller than 64-bit to simulate the situation met in the FHE task. The result in Fig \ref{ablation-percent} shows that ours outperforms the embedded mod in the CUDA toolkit by 8\%.

%% file: sec7-summary.tex
We present \emph{CAT}(Cipher-Acceleration-Textile), a GPU accelerated FHE framework that achieves significant performance gains over existing solutions. Its three-layer architecture, parallel execution strategies, and innovative resource management enable a 2173$\times$ speedup on single operators and considerable acceleration of CKKS, BFV, and BGV schemes.

We also demonstrate its effectiveness in FHE-based Privacy Database Queries, executing complex SQL queries and computations for $10^3$ rows within one second while maintaining memory consumption below 6GB. Extensive testing has been conducted, ensuring its reliability and commercial deployability.

In the future, we plan to optimize bootstrapping procedures, incorporate GPU-accelerated scheme switching mechanisms, and extend the framework's capabilities by integrating additional data query operators. These efforts will further enhance the effectiveness of GPU-accelerated FHE and pave the way for its wider adoption in real-world systems.